\documentclass[11pt]{article}
\usepackage[utf8]{inputenc}
\pdfoutput=1
\usepackage{amssymb,amsmath,mathrsfs,enumerate}
\usepackage{jcappub}
\usepackage{graphicx,rotate,multicol}
\usepackage{float}
\usepackage{tocloft}
\usepackage{subfig}
\usepackage{soul}
\usepackage{multirow}
\usepackage{placeins}
\usepackage{hyperref}
\usepackage{tikz,braket}

\allowdisplaybreaks

\title{\boldmath Constraints on dark matter self-interaction from velocity distribution function in isolated halos}

\author {\bf Tirtha Sankar Ray,$^{a,}$\footnote{tirthasankar.ray@gmail.com} and  Sambo Sarkar$^{b,}$\footnote{sambosarkar92@gmail.com} }
\affiliation[a]{Department of Physics, Indian Institute of Technology Kharagpur, Kharagpur 721302, India}
\affiliation[b]{Laboratory for Symmetry and Structure of the Universe, Department of Physics,
Jeonbuk National University, Jeonju 54896, Korea}

\abstract{Self-interactions facilitate inter-particle redistribution of energy within the dense regions of galactic halos, implying modifications in the density and velocity distribution of dark matter. Simulating dark-matter only isolated halos for a wide range of mass and specific self-scattering cross-sections, we make a systematic study of the impact of self-scattering on the velocity distribution profiles. We report a conservative bound on $\sigma/m$ $\leq 2.7 \rm cm^2/gm$ at $95\%$ C.L. from observations of rotation curves in Milky-Way size galaxies. Sub-leading bounds from LSB galaxy and clusters are also presented.}

\begin{document} 
\maketitle
\flushbottom

\section{Introduction}
\label{sec:intro}

Particulate origin of dark matter has witnessed ever increasing scrutiny. While the microscopic nature of such a dark sector remains elusive, various phenomenological and theoretical consideration provide compelling motivation for such a possibility \cite{Bauer:2017qwy}. This optimism primarily arises because particulate dark matter in addition to its gravitational implications is expected to have interactions with visible matter and possibly within the dark sector. Such interactions may leave imprints in astrophysical and cosmological observables, leading to a rich observable phenomenology of the dark sector \cite{Catena:2016ckl,Egana-Ugrinovic:2021gnu}.

The standard $\Lambda$-Cold Dark Matter ($\Lambda$CDM) cosmological framework \cite{1984Natur.311..517B} together with the standard model (SM) of particle physics has been extremely successful in explaining the formation and evolution of present day observable universe \cite{Lisanti:2016jxe, Pace:2019vrs, DelPopolo:2002sz}. However at the sub-galaxy scales there exists several mismatch between predictions from $\Lambda$CDM and the corresponding observations \cite{Bernardeau:2001qr}. The core-cusp problem \cite{Flores:1994flo, Burkert:1995bur,BoylanKolchin:2003sf}, diversity issue with rotation curves \cite{Oman:2015om,Kamada:2016euw} and dwarf galaxies that are too-big to fail \cite{BoylanKolchin:2011de}, comprise some of these concerns. While these discrepancies may be artifacts resulting from the lack of sophistication in present day $N$-body simulations or proper inclusion of baryonic effects, they may also be indicative of specific properties of DM particles like its self-interaction
\cite{Kaplinghat:2013xca,Pollack:2014rja,Jiang:2022aqw,Zhong:2023yzk,Yang:2024tba}. 
It is worth mentioning that non-standard DM-baryon interactions accompanied with supernova effects  \cite{Vogl:2024ack} and cosmic ray interactions \cite{Martin-Alvarez:2022tfw} have been considered to address the small-scale issues. 

Dark matter self-interactions \cite{Spergel:1999mh, Huo:2019yhk, Kamada:2020buc}, can thermalise the collapsing galactic center, driving the formation of galactic cores \cite{Rocha:2012jg}. 
The idea has been extensively explored in constraining self-interacting DM (SIDM) \cite{10.1046/j.1365-8711.2001.04477.x,Ko:2014nha,Robertson:2020pxj,Rocha:2012jg,Randall:2007ph,Peter:2012jh,10.1093/mnrasl/sls053,Elbert:2014bma,Banerjee:2019bjp}. Within this core the self-interaction is expected to modify the  velocity distribution of DM particles. Naively, an effective rate of DM self-interaction would aide thermalization in the velocity distribution function of DM. In this work we utilize extensive SIDM $N$-body simulations of isolated DM-halos to study its impact on the velocity profile of DM. We investigate the deformation of the velocity profile as a function of galactic radius, halo masses $M_{\rm halo}$ and specific self-scattering cross-section $\sigma/m$. We compare the simulation results with observed rotation curves of galaxies to report constraints on $\sigma/m$ in the range of $2.7-10\, \rm cm^2/gm$.

The paper is organized as follows. In section \ref{sec:Simulate} we discuss the numerical details of the SIDM only $N$-body simulations, and in section \ref{sec:maxwelldist} we discuss the impact of self-scattering on the velocity distribution functions. In section \ref{sec:analytic} and \ref{sec:results} we present our systematic study comparing the simulation results with astrophysical observations to constrain DM self-interactions. Finally we conclude in section \ref{sec:conc}. 
\section{Simulating isolated halos with dark matter self-interaction}
\label{sec:Simulate}

In this article we make a systematic study of the impact of DM self-interaction on the velocity distribution in isolated galactic halos. We simulate dark matter only isolated halos using the $N$-body code \texttt{GADGET} \cite{Springel:2000yr,Volker:2005vol}, modified to include the self-interactions. We follow the algorithm discussed in \cite{Jun:2011jun} and \cite{Ray:2022ydr} to encode the self-interactions. As a particle undergoes scattering with its $n^{th}$ nearest neighbor, weighted by its finite scattering
 probability, the positions and velocities are updated by the Kick-Drift-Kick, leap-frog method \cite{Quinn:1997iy}. Our simulations have been optimized considering a velocity independent, elastic $2 \rightarrow 2$ DM self-interaction. The initial spherically symmetric particle distribution follows the NFW profile \cite{Navarro:1996nav}, and has been generated using the publicly available code \texttt{SphericIC} \cite{Garrison-Kimmel:2013yys,spheric}, which is a modification on the code \texttt{HALOGEN} \cite{Zemp:2007nt,Halogen}. Following the prescription of \cite{Ray:2022ydr}, we extend our simulation set with three different halo masses, evolving them for $10$ Gyrs \cite{Miralda-Escude:2000tvu,Feng:2009hw}. Therefore, extending the work of \cite{Ray:2022ydr}, in this study we perform our analysis with 9 different halo masses. Each DM halo has been simulated with eight values of scattering cross-sections, ranging from $\sigma/m = 0\,\rm cm^2/gm$ (collision-less) to $\sigma/m = 10\,\rm cm^2/gm$. Additionally, for statistical accuracy we have generated and evolved each halo four times, with independent realizations. For each halo, the DM halo mass has been distributed uniformly among $10^6$ particles, within the finite radial extent of the halo, $r_{\rm cut}$. Relevant cosmological parameters have been set from the PLANCK collaboration \cite{Ade:2015xua}. We detail the relevant simulation parameters pertaining to this work in table \ref{tab:halo}, where $r_{\rm so}$ is the initial NFW characteristic radius and $\epsilon$ the initial Plummer equivalent gravitational softening scale \cite{Joshi:1999vf}. Our SIDM halos follow a self-similar evolution in time \cite{Balberg:2002ue,Outmezguine:2022bhq}. Since we perform simulations with specific cross-sections as large as $\sim 10\,\rm cm^2/gm$, we set the time-step for evolution to be small by taking the scattering probability factor to be less than 0.1. This ensures for each particle to have at most one scattering per time-step of evolution \cite{Burkert:2000di}. Details on the stability and convergence of our $N$-body simulations have been discussed in \cite{Ray:2022ydr}. The conditions with which we ran our simulations for generating stable halos i.e. the force-softening, time-step and scattering criterion are consistent with some of the recent works on SIDM simulations \cite{Mace:2024uze,Palubski:2024ibb}.

In the galactic halos, self-interactions become relevant in transferring heat from the periphery to inner dense cores of the galactic halos \cite{Kamada:2019wjo}. This leads to an extended region of thermalised core within a halo whose extent depends on the strength of the self-interaction\cite{Kaplinghat:2015aga}. Consequently, one would also expect self-interactions to impact the velocity of DM within such thermalised regions. A motivated guess would be to observe in this core region a significant deviation of the overall shape and the most probable velocity for the DM distribution, from CDM as self-interactions are introduced \cite{Sean:2017sea}.
\begin{table}[t]
	\centering   
	\bigskip
	\resizebox{15.5cm}{!}{%
	\begin{tabular}{|p{3cm}|p{3cm}|p{2.5cm}|p{2.5cm}|p{2.5cm}|}
		\hline
		Class & $M_{\rm halo}$ ($M_{\odot}$) & $r_{\rm cut}$ (kpc) & $r_{\rm so}$(kpc) & $\epsilon$ (kpc) \\
		
		\hline \hline
		\setlength\arraycolsep{10pt}	
		& $10^{10}$		& 43.7 & 0.43 & 0.02 \\ 
		
		LSB Scale&$5\times 10^{10}$ & 74.3 & 0.74 & 0.02 \\ 
		
		& $10^{11}$			& 93.4 & 0.94 & 0.03 \\ 
		\hline \hline
		
		&$5 \times 10^{11}$ & 158  & 1.58 & 0.06 \\ 
		
		Milky-Way Scale&$10^2$		& 199  & 1.99 & 0.06 \\ 
		
		&$5 \times10^{12}$ & 339  & 3.39 & 0.06 \\ 
		
		\hline \hline
		&$10^{14}$          & 913  & 9.13 & 0.28 \\
		
		Cluster Scale &$5 \times 10^{14}$& 1551 & 15.51& 0.58 \\
		
		&   $10^{15}$          & 1952 & 19.52& 0.58 \\
		\hline 
	\end{tabular}}
	\caption{Parameter set for generating the initial distribution of the isolated halos following \cite{Ray:2022ydr,Jun:2011jun}.}
	\label{tab:halo}
\end{table}
\section{Velocity distribution  of Dark Matter}
\label{sec:maxwelldist}
\begin{figure*}[t]
	\begin{center}
		\subfloat[\label{sf:chisqfit1}]{\includegraphics[scale=0.25]{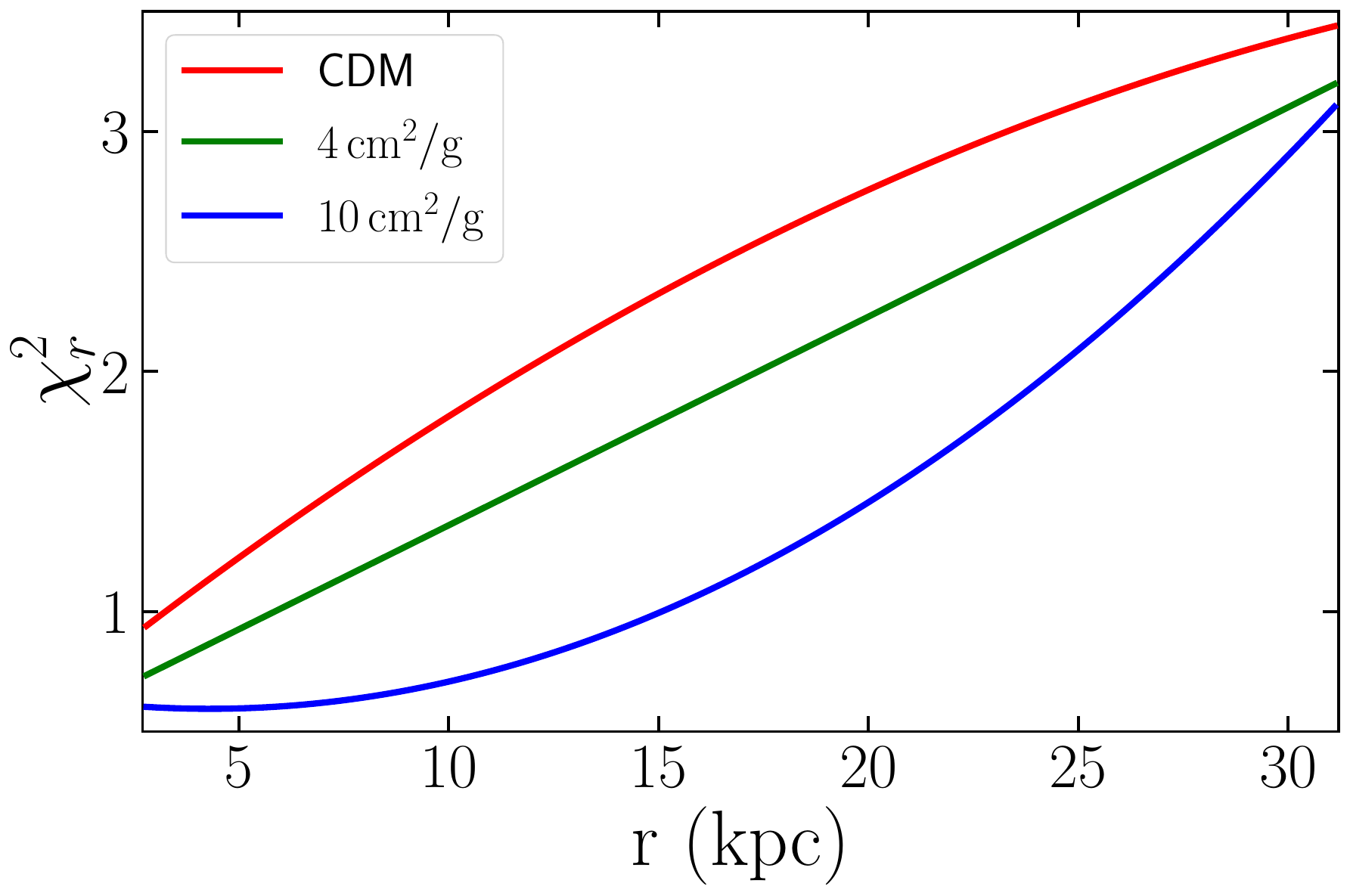}}
		\subfloat[\label{sf:chisqfit2}]{\includegraphics[scale=0.25]{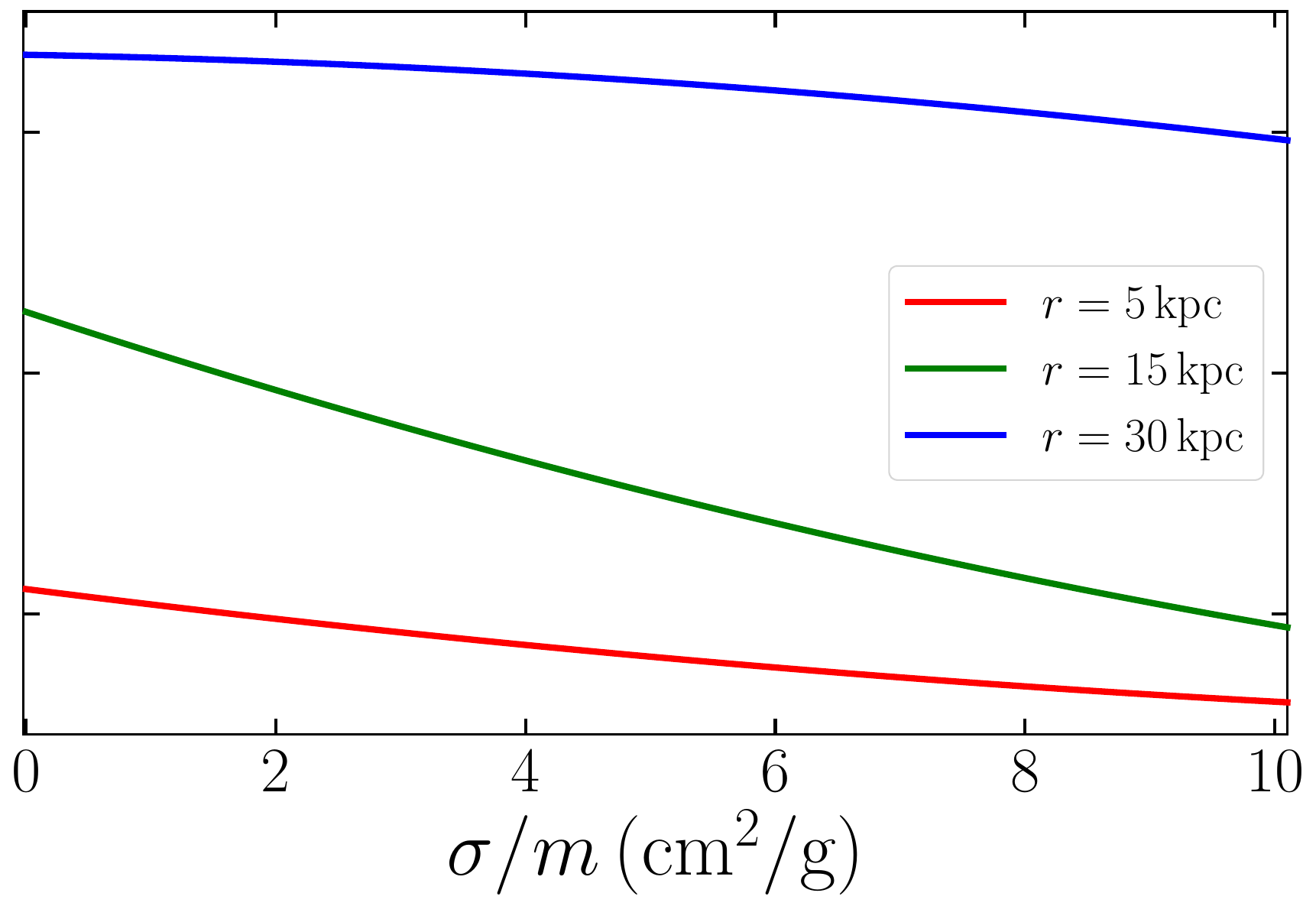}}
		\caption{Reduced $\chi_{\rm r}^2$ as a function of the radial distance from the center (left) and $\sigma/m$ (right), for the MB distribution for DM halo mass of $10^{12}\rm M_{\odot}$.}
		\label{fig:chisqfit}
	\end{center}
\end{figure*}

For a structured discussion on the impact of dark matter self-interaction on their velocity distribution function  inside halos, an identification of some distribution profile are now in order. The CDM velocity distribution once virialized in the halos are not expected to exhibit any thermalized behavior. It is usually assumed that the velocity distribution function follows the Maxwell-Boltzmann (MB) distribution with appropriate cut-offs, $v_{\rm esc}$ \cite{Binney:1987bin}. The thermal behavior of DM, across different mass scales has been studied in  \cite{Munari:2013mh,Zhang:2024hrq,Palubski:2024ibb,Yang:2024uqb}. We consider both the MB and non-thermal distributions with pure statistical motivation for comparisons:
\begin{itemize}

\item \textbf{Maxwell-Boltzmann distribution:} 
A popular assumption for the velocity distribution of DM inside galactic halos  is the  isotropic and isothermal MB function. In order to account for the finite extent of a halo, a cut-off velocity given by the escape velocity of particles is introduced. This along with a CDM like radial distribution constitute the standard halo model (SHM) \cite{LyndenBell:1966bi, 1990ApJ...353..486L}, and can be expressed as:
\begin{equation}
f(\mathbf{v})=
\begin{cases}
\frac{1}{N}\left[\exp{\left(-\frac{|\mathbf{v}|^{2}}{v_{0}^{2}}\right)}\right] &  |\mathbf{v}| \leq v_{\rm esc} \\
0 & |\mathbf{v}| > v_{\rm esc},
\end{cases}
\label{eq:MB}
\end{equation}
where $N$ denotes a normalization constant, $v_0$ is the measure of its most probable velocity and $v_{\rm esc}$ sets the maximum allowed DM velocity for a given galaxy.

\item \textbf{Gaussian distribution:} In the absence of an underlying mechanism driving  DM thermalisation, the often quoted MB distribution for halos remains difficult to motivate. From a purely statistical perspective one may adopt the more generalized normal distribution for velocities about a mean value \cite{BeraldoeSilva:2013efk}, and can be represented as:
\begin{equation}
f(\mathbf{v})=
\begin{cases}
\frac{1}{N}\left[\exp{\left(-\frac{|\mathbf{v}-\mu|^{2}}{2\sigma^{2}}\right)}\right] &  |\mathbf{v}| \leq v_{\rm esc} \\
0 & |\mathbf{v}| > v_{\rm esc},
\end{cases}
\label{eq:ND}
\end{equation}
where $N$ denotes the normalization constant, $\sigma$ is the dispersion, $\mu$ is the mean velocity of the DM particles and the finite velocity cutt-off is set by $v_{\rm esc}$ \cite{Binney:1987bin}.   
\begin{figure*}[t]
	\begin{center}
		\subfloat[\label{sf:Xerest1}]{\includegraphics[scale=0.191]{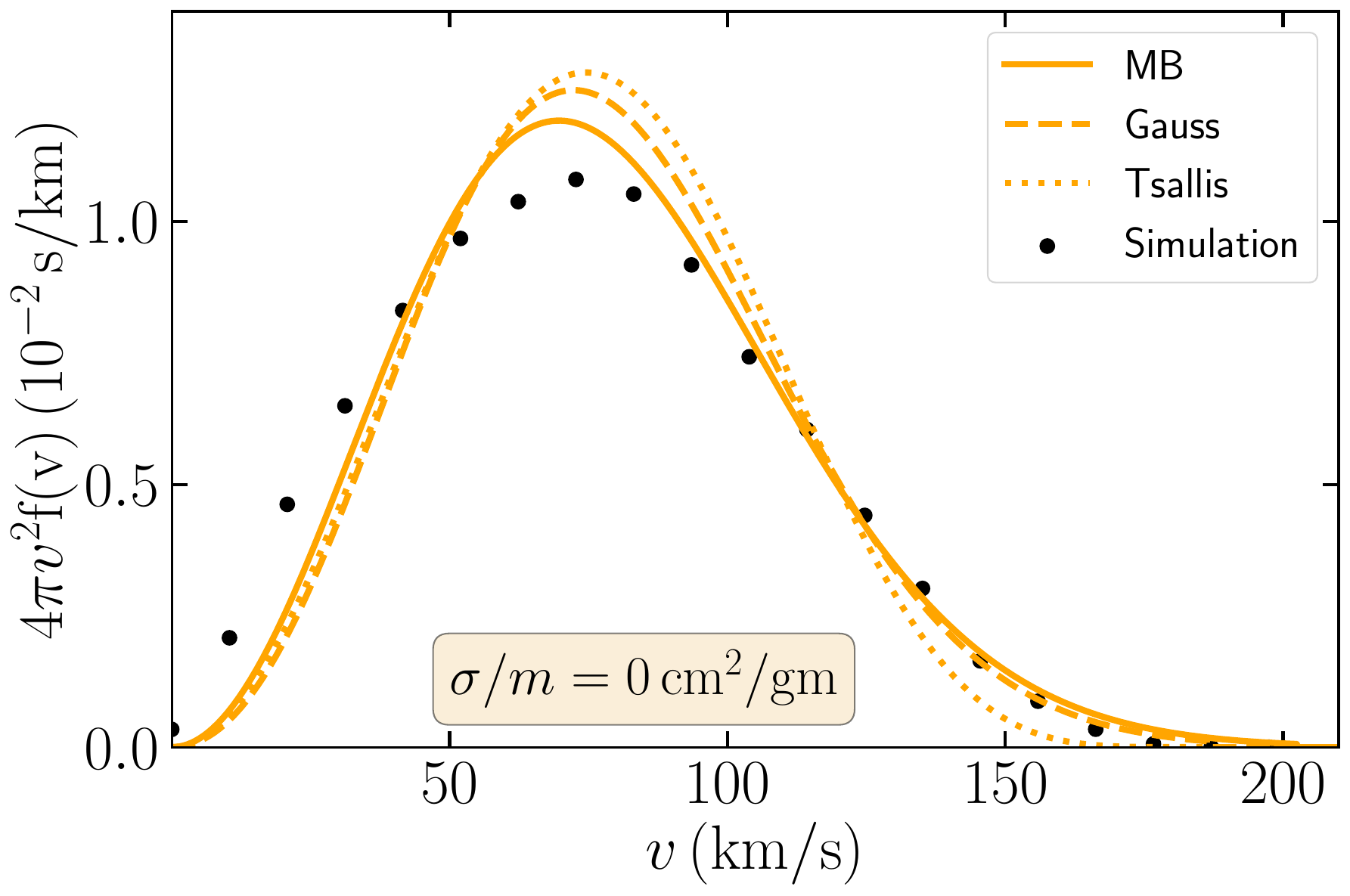}}
		\subfloat[\label{sf:Sirest1}]{\includegraphics[scale=0.191]{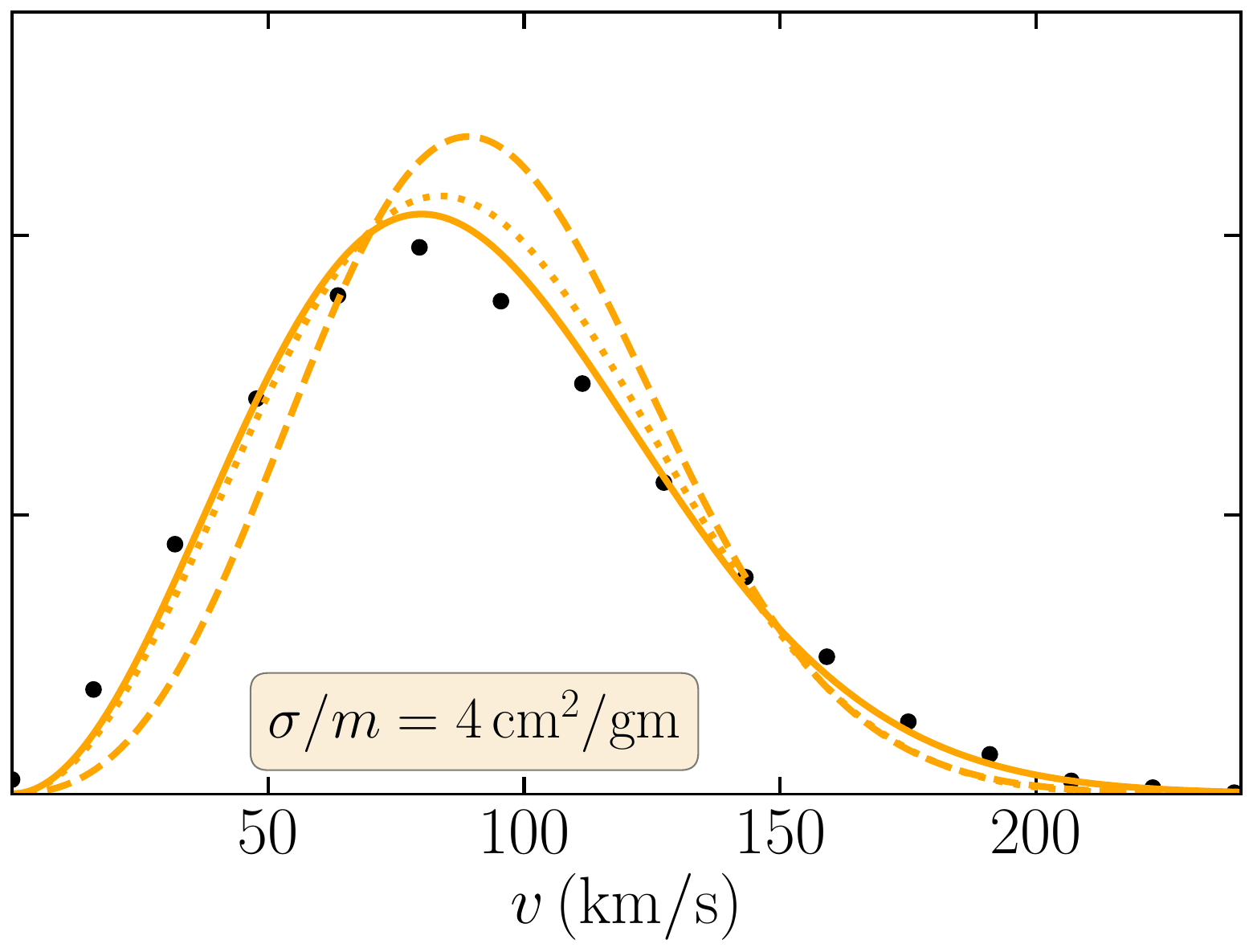}}
		\subfloat[\label{sf:Gerest1}]{\includegraphics[scale=0.191]{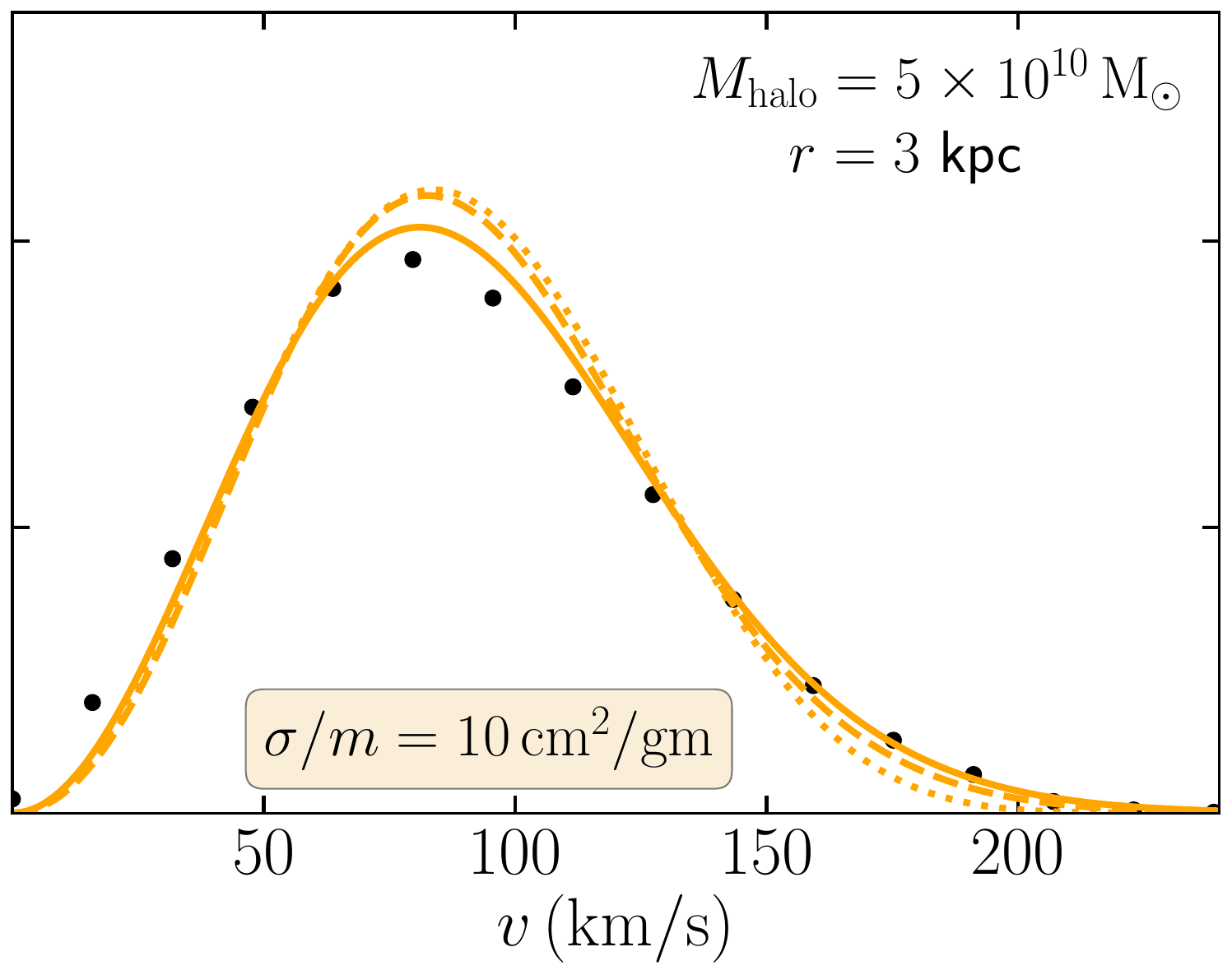}}\\
		\subfloat[\label{sf:Xerest1}]{\includegraphics[scale=0.191]{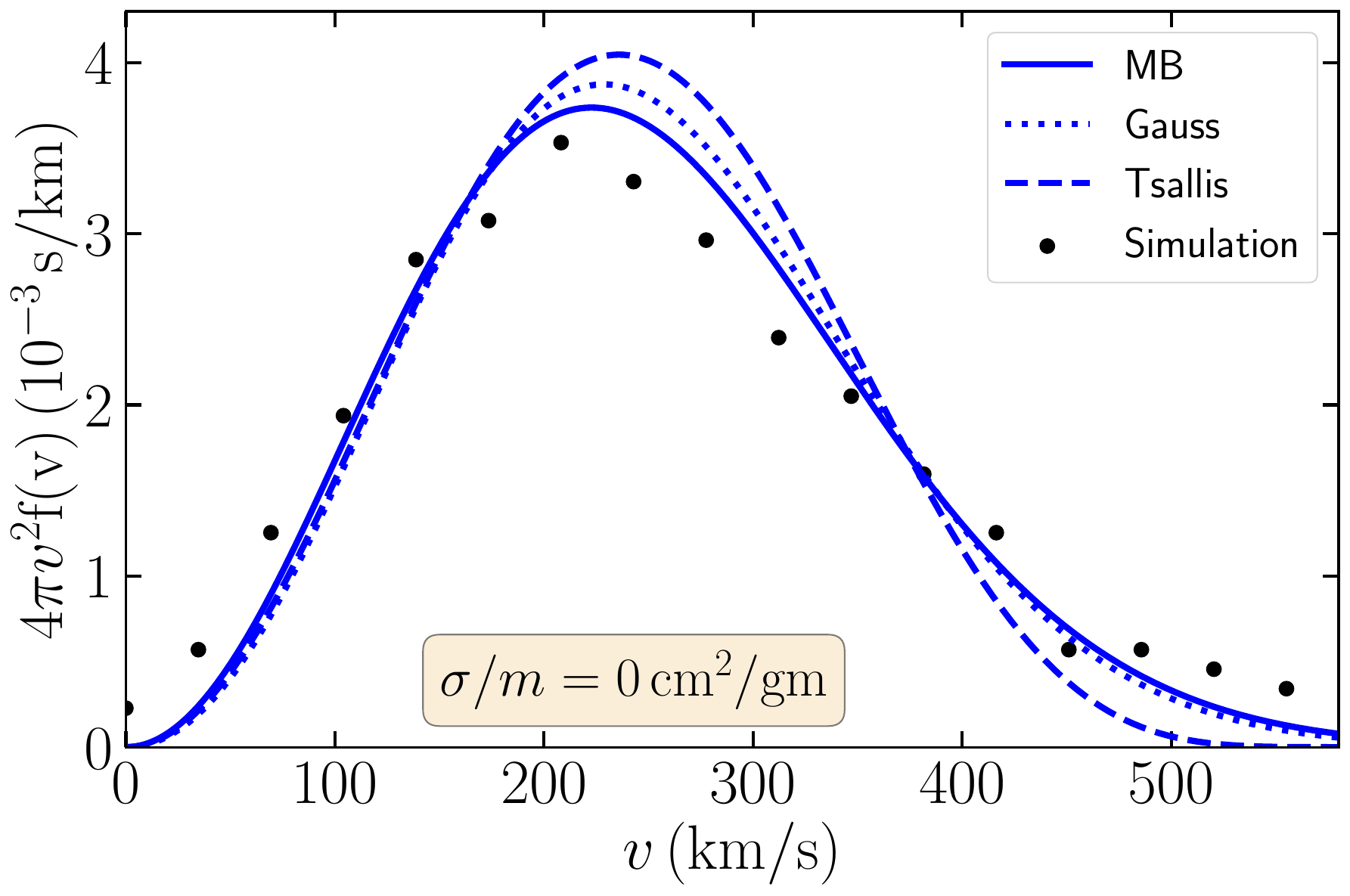}}
		\subfloat[\label{sf:Sirest1}]{\includegraphics[scale=0.191]{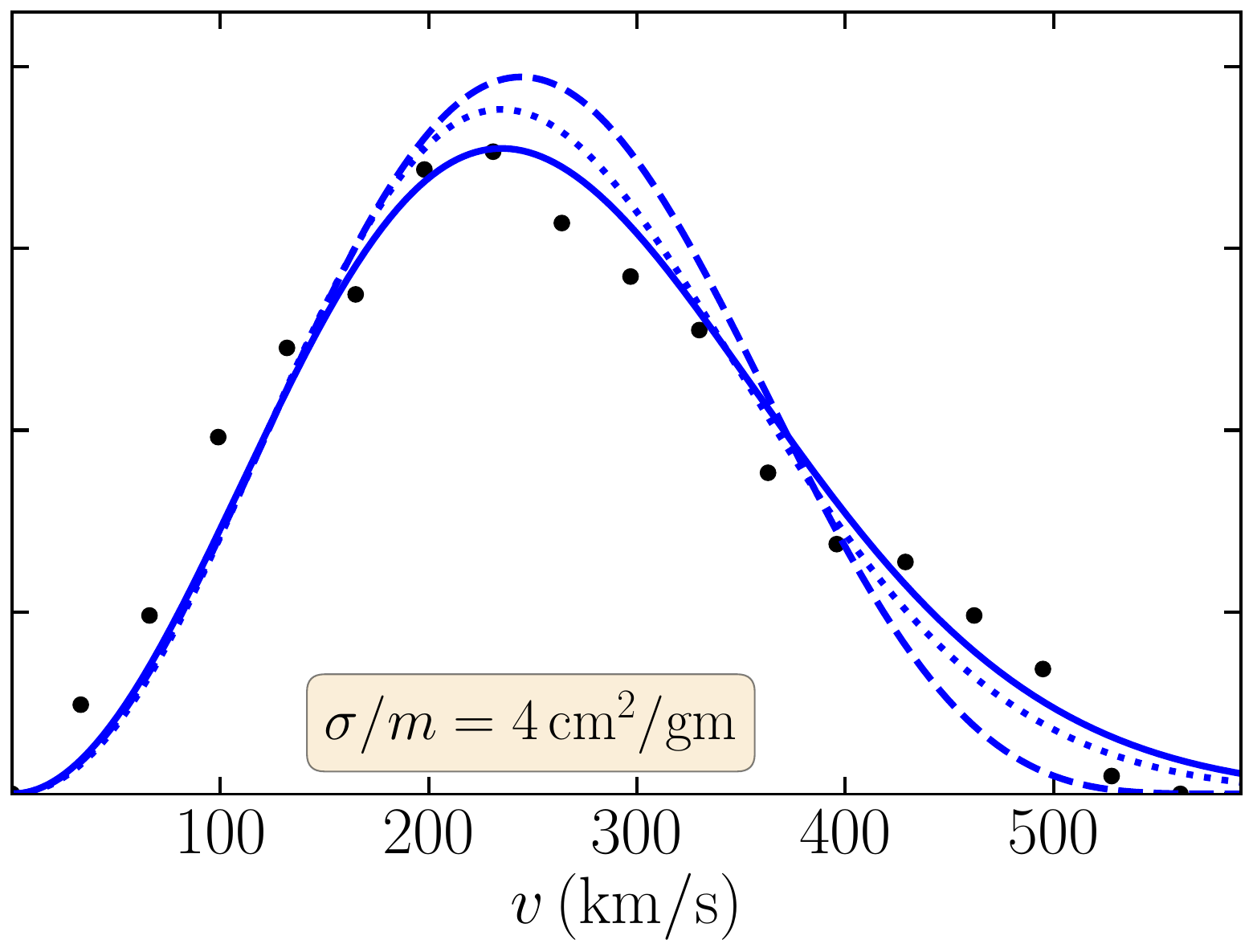}}
		\subfloat[\label{sf:Gerest1}]{\includegraphics[scale=0.191]{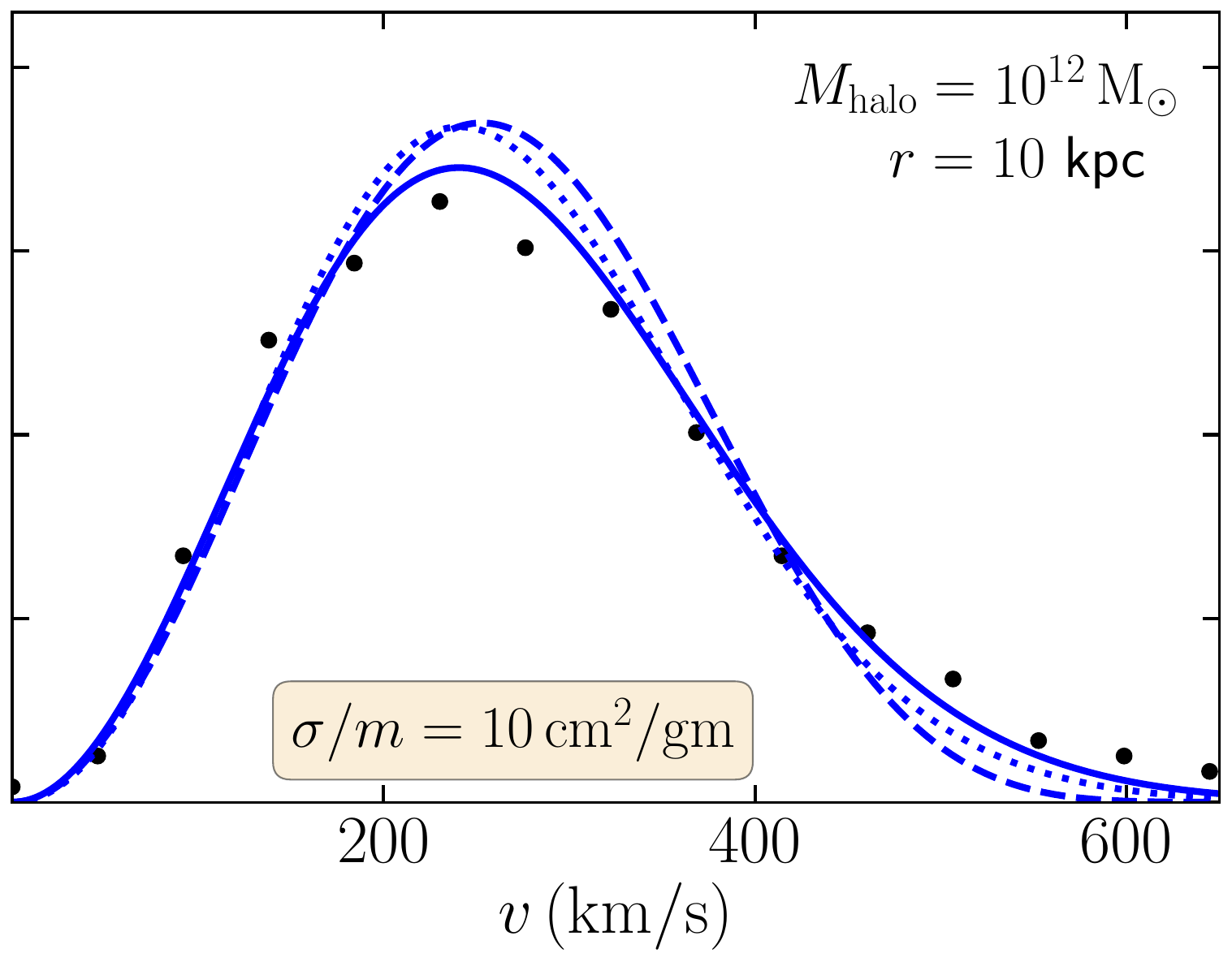}}\\
		\subfloat[\label{sf:Xerest1}]{\includegraphics[scale=0.191]{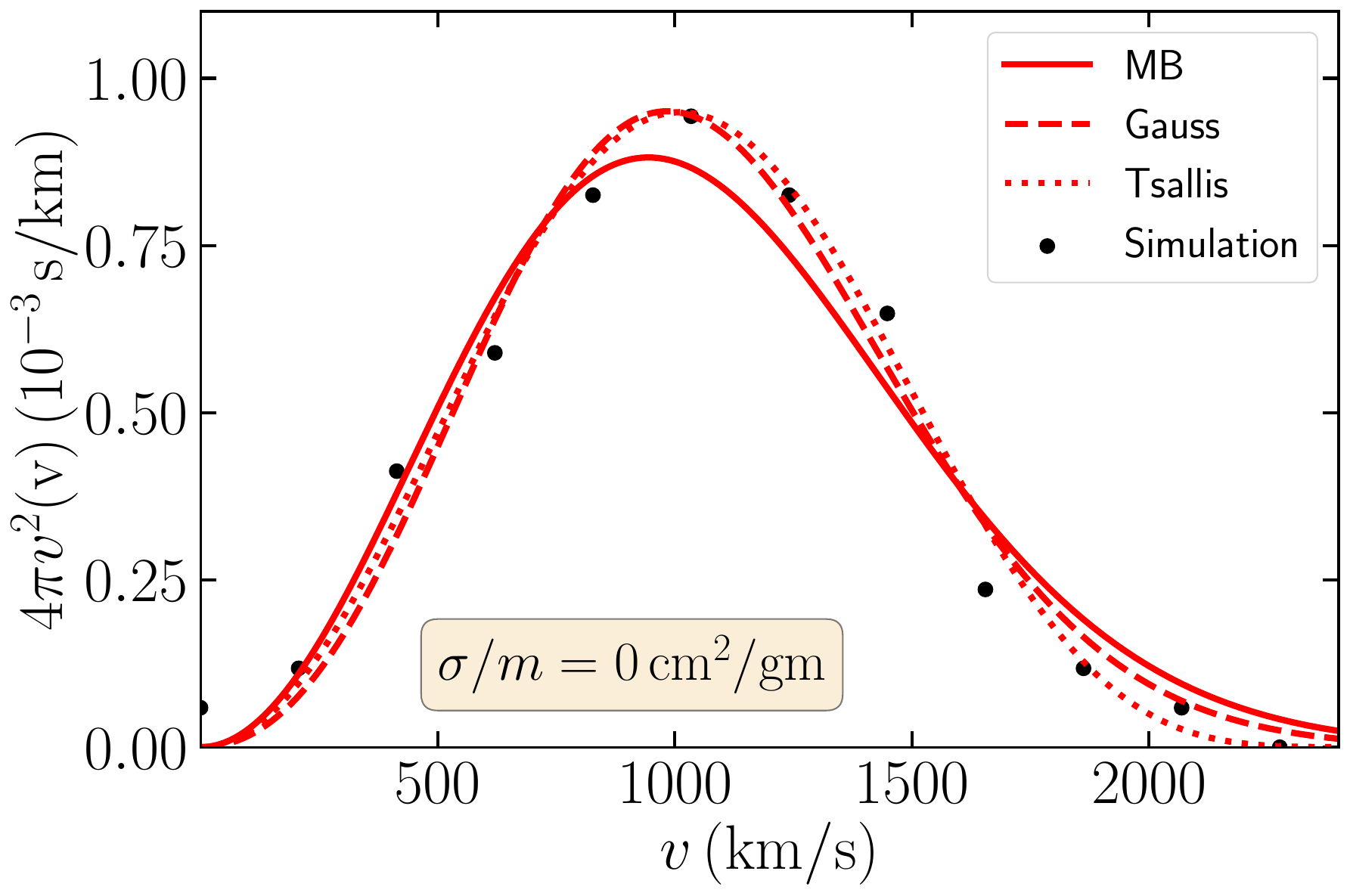}}
		\subfloat[\label{sf:Sirest1}]{\includegraphics[scale=0.191]{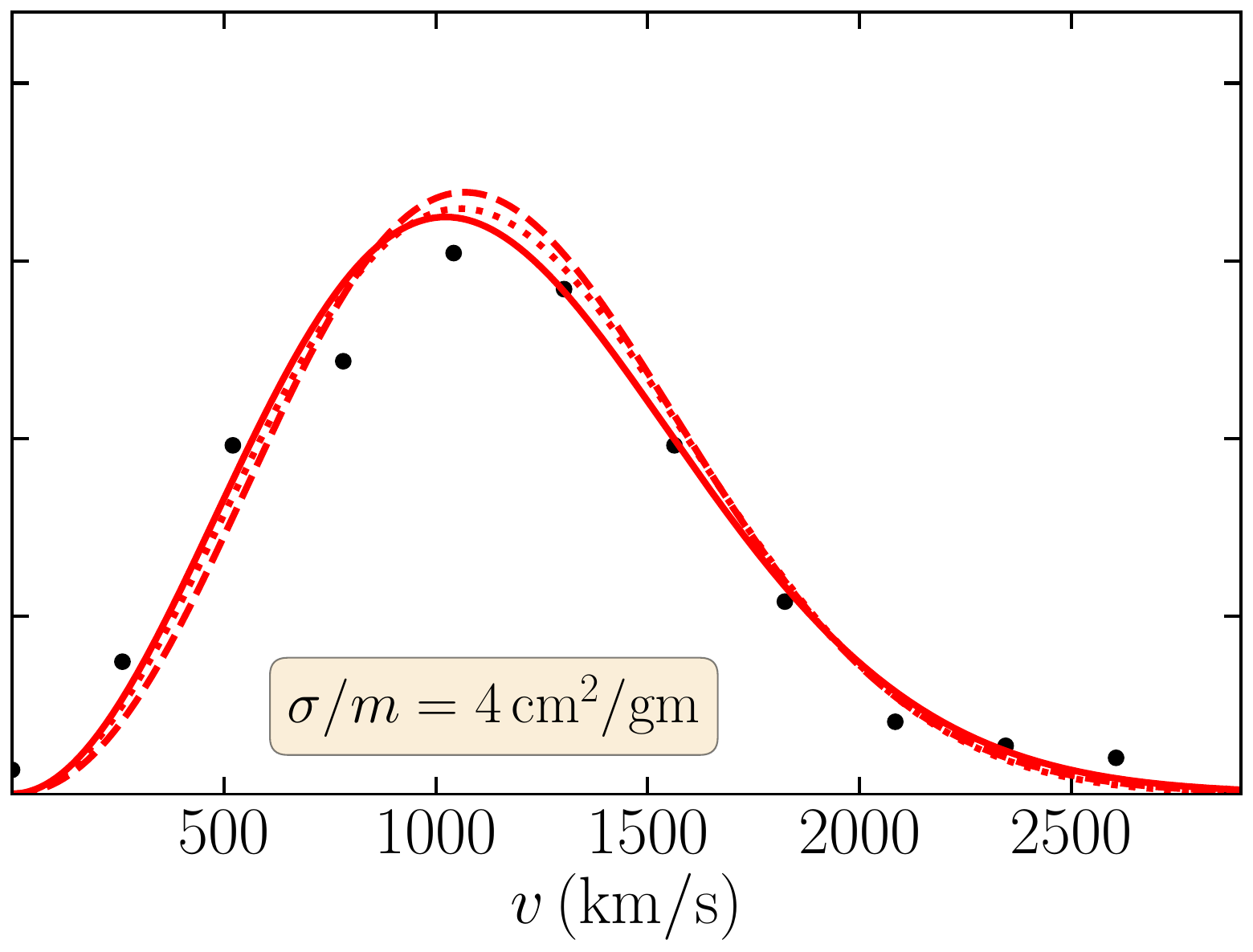}}
		\subfloat[\label{sf:Gerest1}]{\includegraphics[scale=0.191]{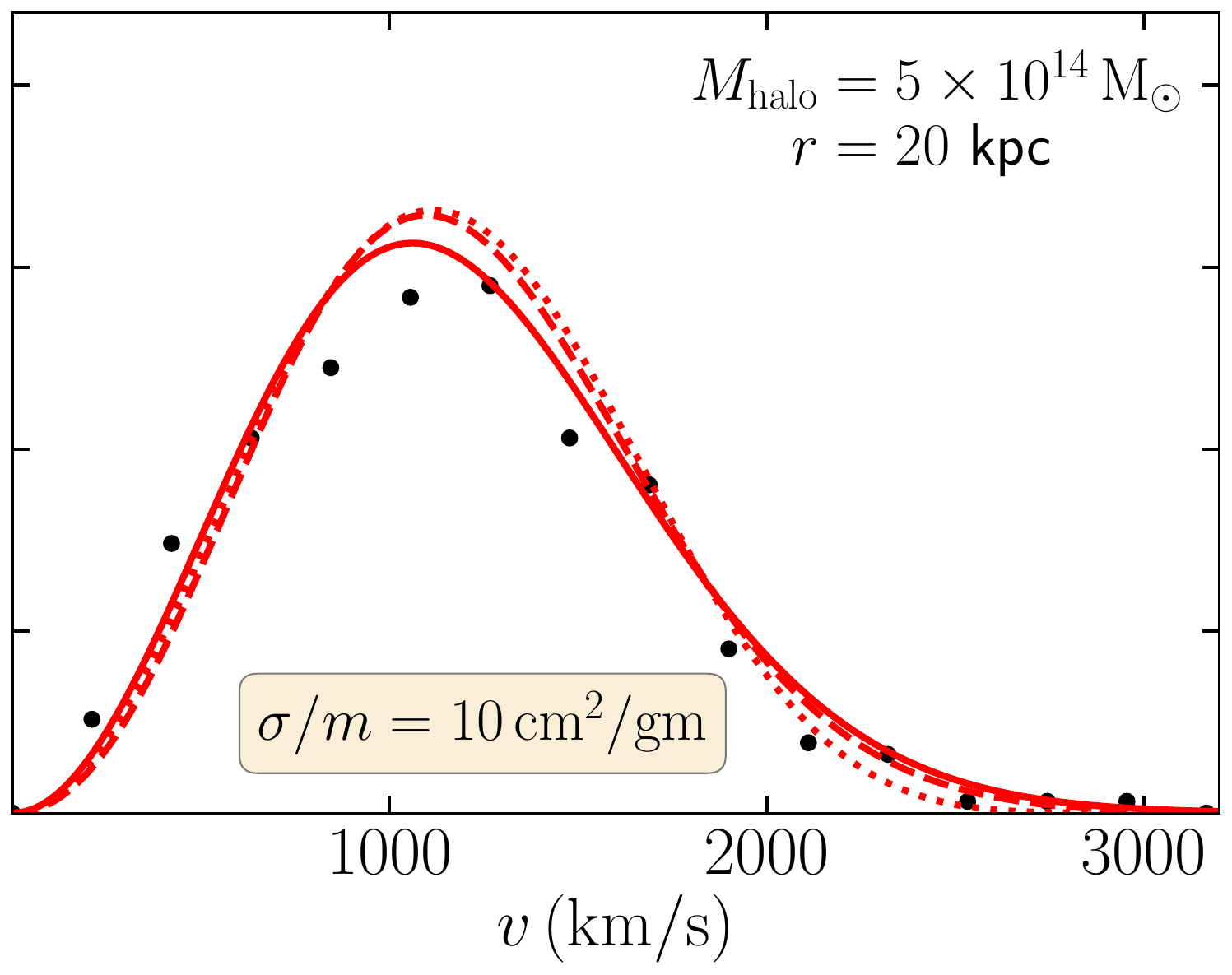}}
		\caption{Fits to our simulated DM velocity distribution (black points), for three benchmark halo mass(radius): $5\times10^{10}M_{\odot}$($3$ kpc), $10^{12}M_{\odot}$($10$ kpc) and $5\times10^{14}M_{\odot}$($20$ kpc), shown in the top, middle and bottom panels respectively. The solid, dashed and dotted curves represent the best fits to the Maxwell-Boltzmann, Gaussian and Tsallis distributions respectively. The left, middle and right panels are for $\sigma/m$ $0,\,4$ and $10 \, \rm cm^2/gm $ respectively.}
		\label{fig:distfit-2}
	\end{center}
\end{figure*}
\item \textbf{Tsallis distribution:} Obtained from a factorization approximation of the non-extensive Tsallis statistics \cite{Tsallis:1987eu}, the Tsallis distribution is  a generalization of the Boltzmann-Gibbs entropy. It is widely used in high energy collisions, Bose-Einstein condensation \cite{MILLER2006357}, neutron star EOS \cite{Menezes:2014wqa}, early universe cosmology \cite{szczniak2018nonparametric}, etc. The velocity distribution function is given by
\begin{equation}
f(\mathbf{v})=
\begin{cases}
\frac{1}{N}\left[1-\left(1-q\right)\frac{|\mathbf{v}|^{2}}{v_{0}^{2}}\right]^{\frac{1}{1-q}} &  |\mathbf{v}| \leq v_{\rm esc} \\
0 &  |\mathbf{v}| > v_{\rm esc},
\end{cases}
\label{eq:Tsallis}
\end{equation}
here the symbols carry their usual meaning. With $q<1$, the escape velocity is determined by the relation $v_{\rm esc}^2=v_0^2/(1-q)$ making the distribution appealing with an  implicit cut-off. However, for $q>1$  the escape velocity remains undefined, and for the $q \to 1$ limit, the Tsallis distribution reduces to the Gaussian profile \cite{Vogelsberger:2008qb, Kuhlen:2009vh, Ling:2009eh}.
\end{itemize}
Apart from the MB distribution and other empirical models, simulations often use a self-consistent formalism known as the Eddington inversion mechanism \cite{Lacroix:2020lhn}, in order to derive the DM phase-space distribution of halos. This however often requires the assumption of an underlined symmetry of the system. Several other distributions having empirical motivations are often used in modeling the velocity of DM i.e. Double power law\cite{Lisanti:2010qx}, King profile \cite{2008gady.book.....B}, Non-extensive Tsalis statistics \cite{Tsallis:1987eu}, Mao et al. \cite{Mao:2013nda}, etc. Many of which have shown significant effects on both direct and indirect searches of DM \cite{Maity:2020wic,Bose:2022ola}. However, we will not consider these possibilities in order to leave the discussion tractable. It can be argued that DM self-interactions within the dark sector might result in Maxwellian velocities by means of thermalisation, previously absent in CDM halos \cite{Vogel:2013vol}. In figure \ref{fig:distfit-2}, we fit\footnote{We fit the simulated DM velocity distributions to the Maxwell-Boltzmann distribution and obtain the best-fit values of $v_0$, using the inbuilt python function \texttt{scipy} \cite{Virtanen:2019joe}.} the three distributions outlined above to a set of our simulated data for three values of $M_{\rm halo}$ and $\sigma/m$ at specific radial points, broadly within the core region. To quantitatively compare the distributions as candidates for the velocity profile of SIDM halos, we obtain reduced chi-square $\chi_{\rm r}^2$ as a function of the specific DM self-interaction. For a given model of velocity profile we use the maximum simulation error from the simulated halos. In figure \ref{fig:chisqfit}, the reduced chi-square $\chi_{\rm r}^2$ variations for different radial positions and scattering cross-sections reveal a gradual decreases in $\chi_{\rm r}^2$, as we go towards higher values of $\sigma/m$, for the MB distribution. This can be construed as a consequence of self-interaction driven thermalization, specially within the core region where it aids the transfer of heat from the periphery of a halo. Expectedly, the fit to thermal distributions become worse as we move away from the central core region as evident from figure \ref{fig:chisqfit}. 

\begin{figure*}[t]
	\begin{center}
		\subfloat[\label{sf:Xerest1}]{\includegraphics[scale=0.18]{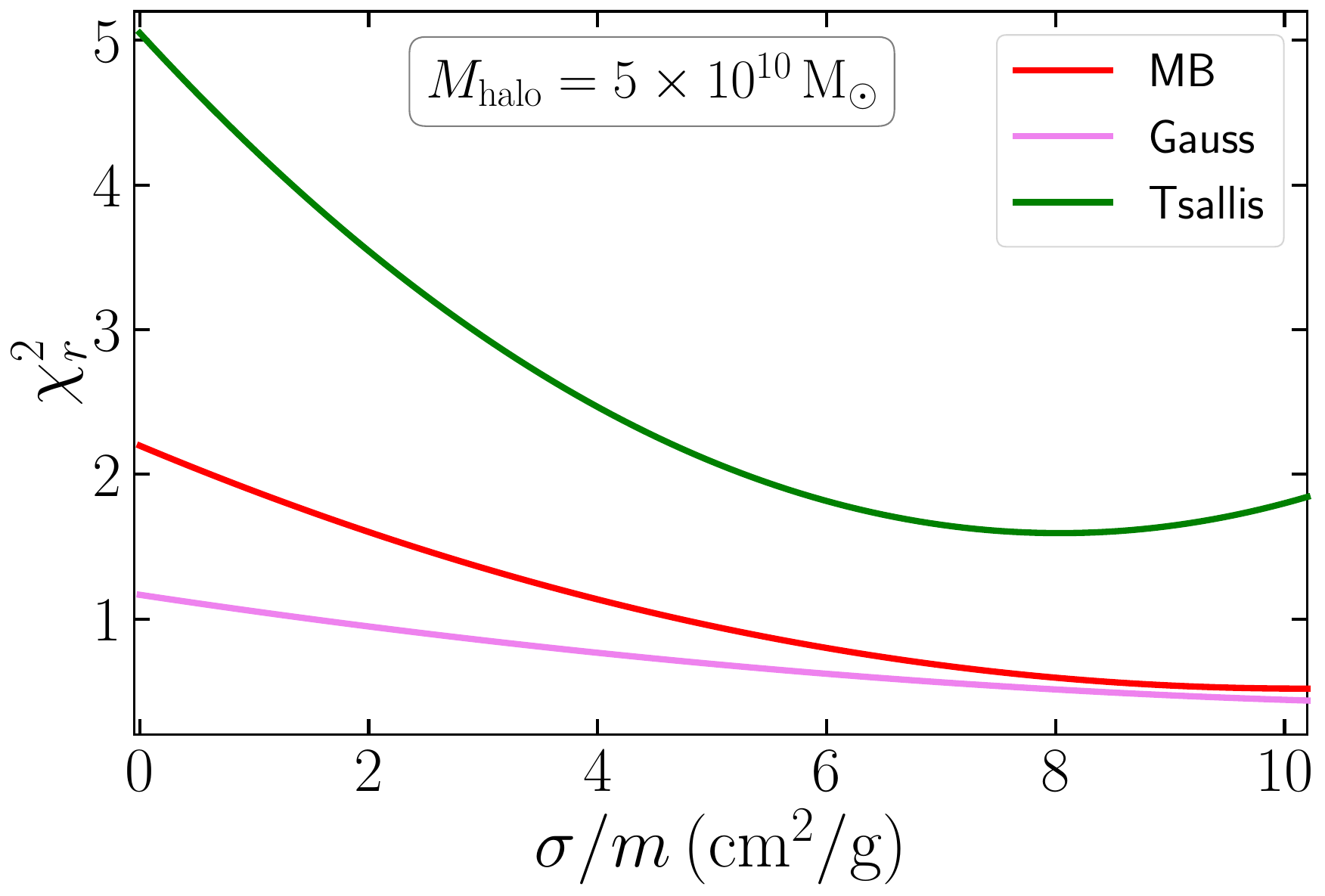}}
		\subfloat[\label{sf:Sirest1}]{\includegraphics[scale=0.18]{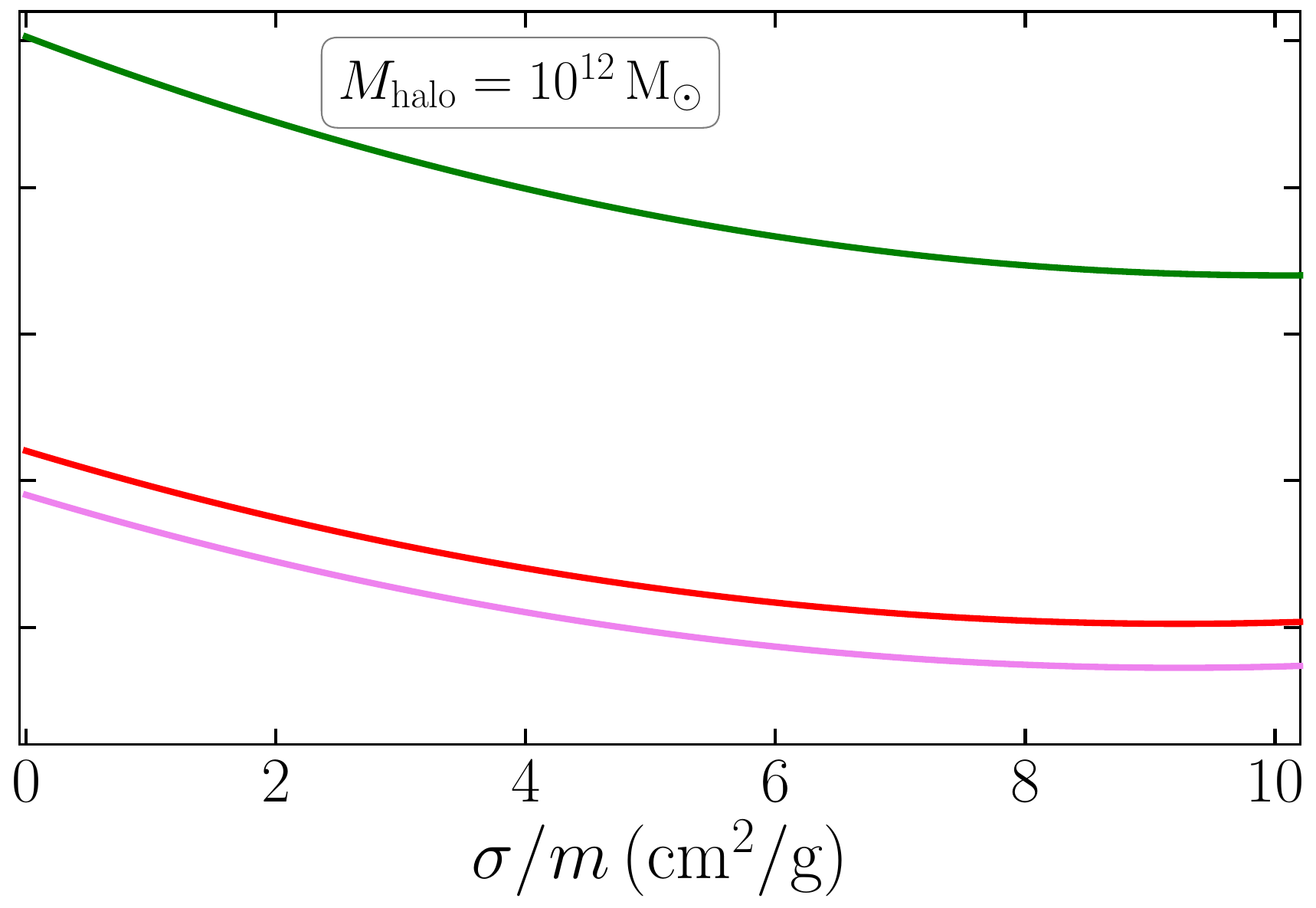}}
		\subfloat[\label{sf:Gerest1}]{\includegraphics[scale=0.18]{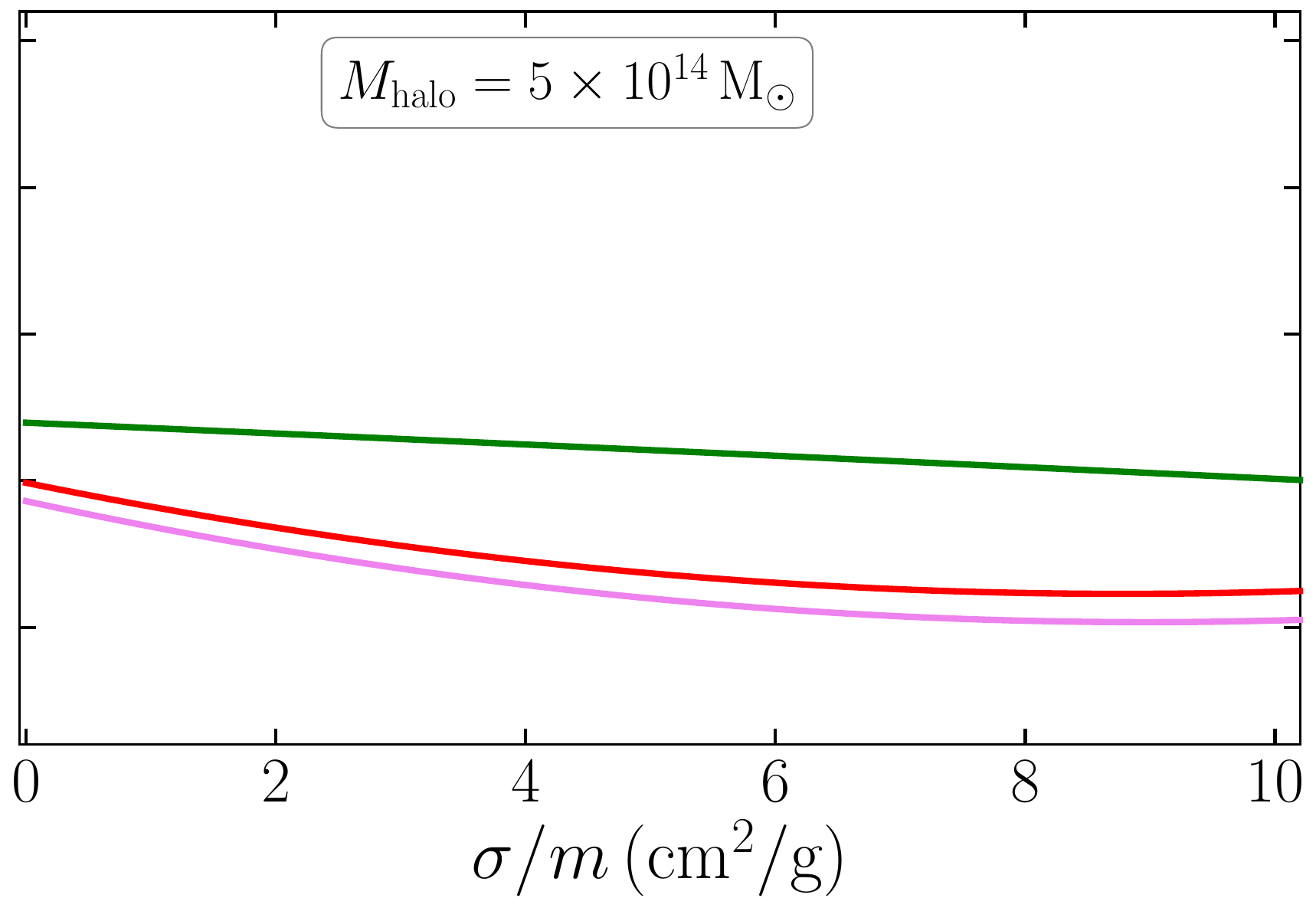}}
		\caption{$\chi_{\rm r}^2$ as a function of $\sigma/m$, for the three distributions. Left, middle and right panels are for DM halo mass(radius) of $5\times10^{10}M_{\odot}$($3$ kpc), $10^{12}M_{\odot}$($10$ kpc) and $5\times10^{14}M_{\odot}$($20$ kpc) respectively.}
		\label{fig:distcomp}
	\end{center}
\end{figure*}

A quantitative comparison between the analytical distributions considered in this section is presented in figure \ref{fig:distcomp}, where we compare the goodness of fit as a function self-interaction cross-section for the three distribution functions. This procedure is repeated for the three class of halos considered viz. LSBs, Milky Way sized galaxies and galaxy clusters. The plots are made at radial values of the halos that lie within thermalisation regions where self-interactions are active. As can be easily seen from the plots, the generalized Gaussian profile gives marginally better fit across the mass range. However, given the relative error between the normal distribution and the often utilized MB distribution function not being catastrophic, for rest of the analysis we will utilize the latter to constraint the self-interaction of DM by comparing the SIDM distribution function fits with observed galactic rotation curves.
\section{Generalized MB distribution in SIDM}
\label{sec:analytic}

The distortions of velocity distribution profiles in SIDM halos dependent on both the specific self-interaction cross section as well as the halo size, as can be seen from figure \ref{fig:distcomp}. We will compare this distortion introduced in the distribution functions with observations of galactic rotation curves to constraint the DM self-scattering cross section.

To capture the leading effect of the self-interactions within the MB frame, we generalize the most probable velocity $v_0$ at a given radial distance in a halo, as defined in Eq. \ref{eq:MB}, to be a generic function  of the halo mass ($M_{\rm halo}$) and the specific self-scattering cross-section ($\sigma/m$). We assume the other parameter in the MB distribution function give by $v_{\rm esc}$ to remain fixed at its value set by the total halo mass. The dependence of the most probable velocities $v_0,$ at a given radius of the galaxy, on the DM halo properties can be expressed utilizing some empirical ansatz as given below.
\begin{enumerate}
	\item \textbf{Polynomial expansion (P1)}:
	Consider a polynomial expansion  for the most probable velocity on the halo mass and the specific self-interaction cross-section\footnote{In this expression we ignore the cross-term owing to its limited statistical impact on the goodness to fit discussed later in the article.}.
	\begin{equation}
	\label{eq:poly_fit}
	v_{\rm o}(\frac{\sigma}{m},M_{\rm halo}) = a_0 + \sum_{i=1}^{n} a_i  \left(\frac{\sigma/m}{cm^2/gm} \right)^i + \sum_{i=1}^{n} b_i \left(\frac{M_{\rm halo}}{10^{10}M_{\odot}} \right)^i 
	\end{equation} 
	Keeping leading terms up to order two $(n=2)$ with undefined coefficients  $a_0,a_1,a_2,b_1,b_2$ suffices to capture the leading effect of self-interaction in modulating the MB statistics.
	\item \textbf{Power-law expansion (P2)}:
	An alternative approach is to consider a power-law dependence over the two parameters  $\sigma/m$ and $M_{\rm halo}$ as given by,
	\begin{equation}
	\label{eq:pow_fit}
	v_{\rm o}(\frac{\sigma}{m},M_{\rm halo}) = a_0 + a \left(\frac{\sigma/m}{cm^2/gm} \right)^{\alpha} + b \left(\frac{M_{\rm halo}}{10^{10}M_{\odot}} \right)^{\beta},
	\end{equation} 
	here $a_0,a,b$ are the coefficients and $\alpha$ and $\beta$ are the power-law index. 
\end{enumerate}
The task now is to evaluate the coefficients of the profiles defined in Eq. \ref{eq:poly_fit} (P1) and Eq. \ref{eq:pow_fit} (P2), by fitting them to the SIDM $N$-body simulation data. With $v_{o}^{sim}$ obtained from $N$-body simulations of isolated halos, we use Markov Chain Monte Carlo (MCMC) \cite{Speagle:2019ffr} affine invariant sampler \texttt{emcee}\cite{2013PASP..125..306F} to fit the parameters for the polynomial and power-law profiles defined above. We set the range for priors using circular velocities at a given galactocentric radius, for LSB and the spiral galaxies. For galaxy-clusters we use the line-of-sight velocity dispersion measurements from Brightest Cluster Galaxies (BCG). Next we systematically use this method for the Milky-way(MW) like spiral galaxy, F563-1 like  LSB and the A611 like galaxy cluster, to fit the profiles of $v_0$ as defined above.
\section{Limits on SIDM from modified velocity distribution function}
\label{sec:results}
In this section we compare the fitted $P1$ and $P2$ profiles for various sized isolated halos with the observed rotational curve data from specific galaxies or clusters. This comparison allows us to put competitive constraints on the specific self-scattering cross-section of dark matter. 
\subsection{Milky-Way like spiral galaxies}
\label{subsec:MW}
\begin{figure*}[t]
	\begin{center}
		\subfloat[\label{sf:mwpolymargin1}]{\includegraphics[scale=0.267]{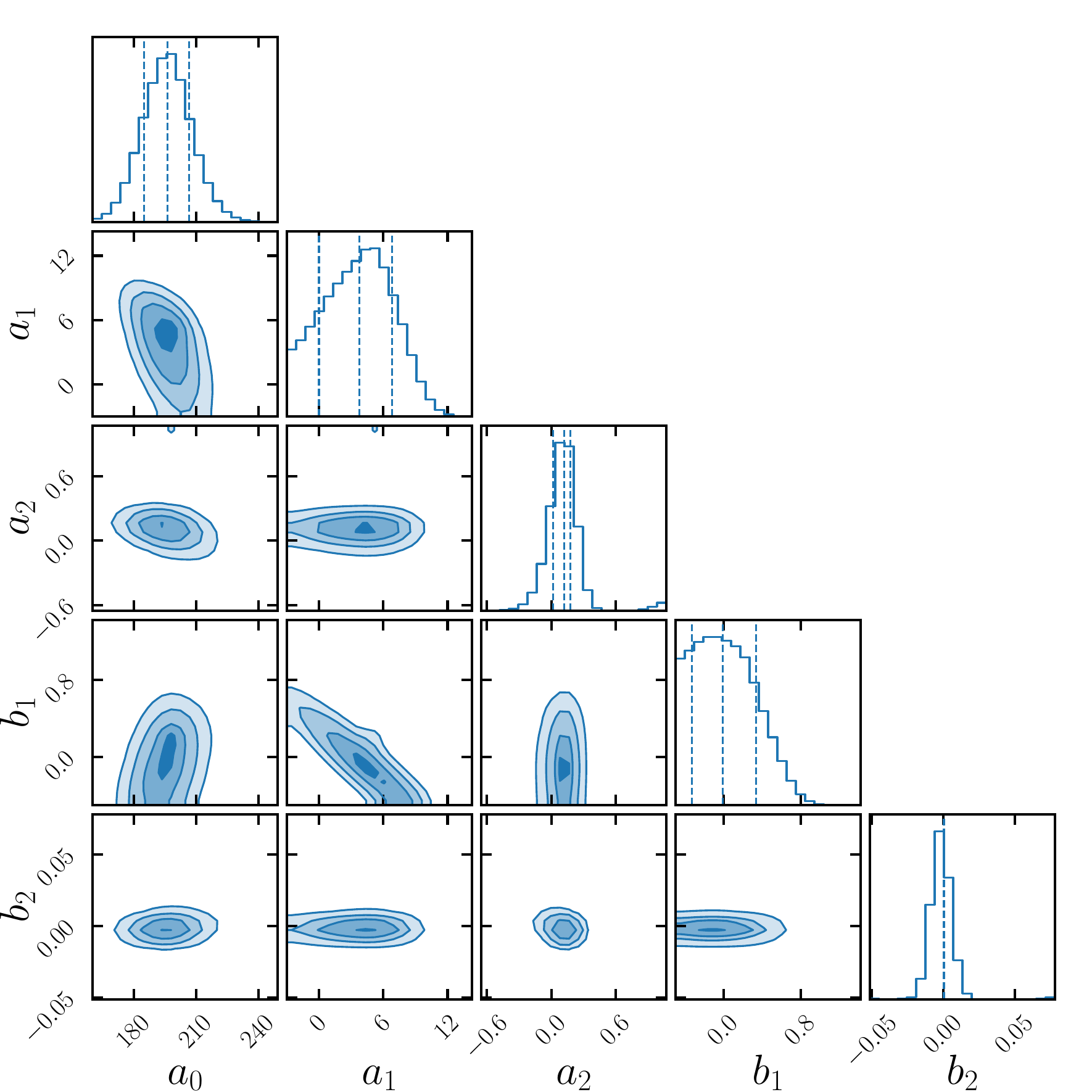}}
		\subfloat[\label{sf:mwpolymargin2}]{\includegraphics[scale=0.267]{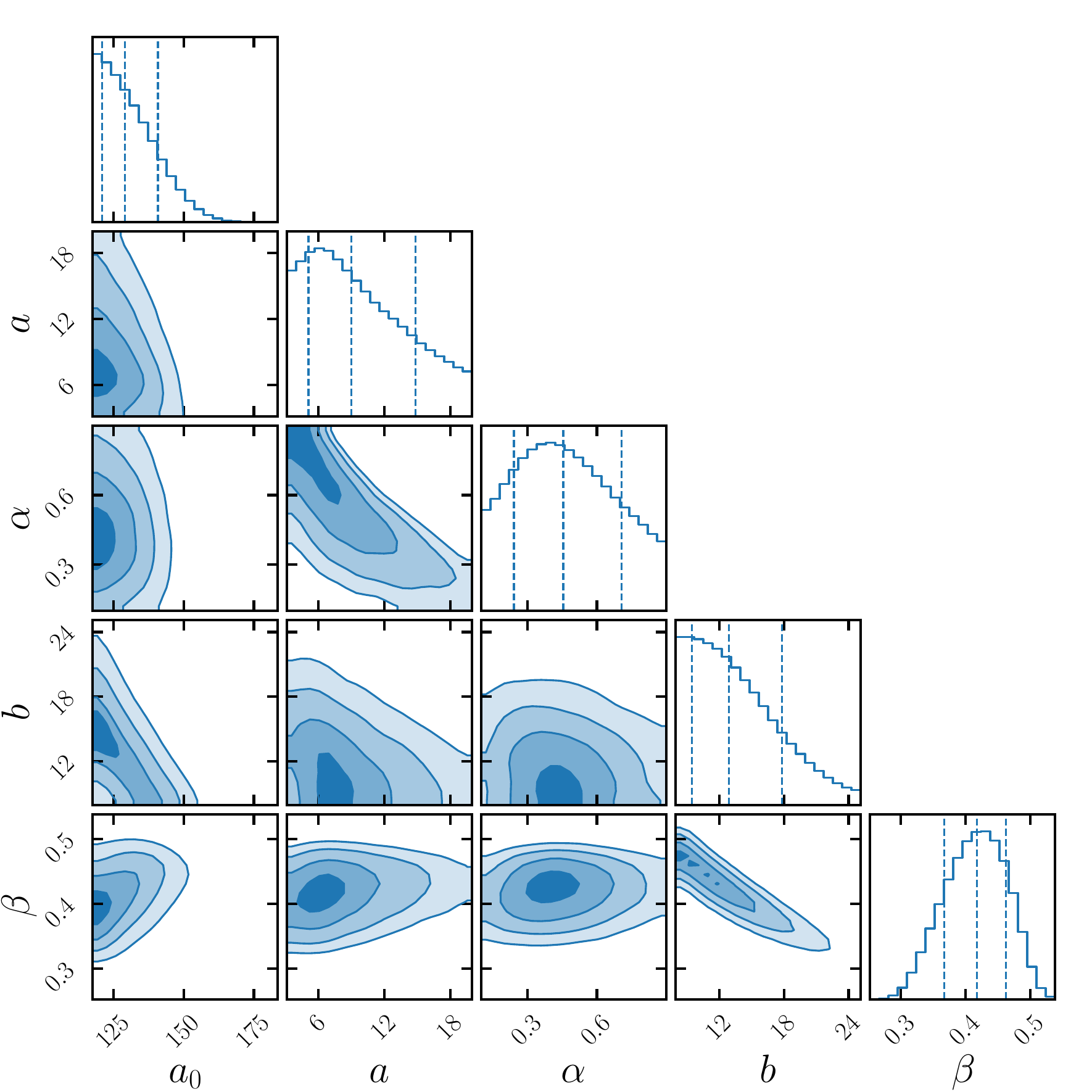}}
		\caption{Posterior PDF for halos in the considered MW mass range. Initialized for circular velocity at a galactocentric radius of 8.2 kpc. The polynomial and power-law functions are represented in the left and right panels respectively.}
		\label{fig:mwpdf1}
	\end{center}
\end{figure*}

\begin{figure*}[t]
	\begin{center}
		\subfloat[\label{sf:Xerest1}]{\includegraphics[scale=0.191]{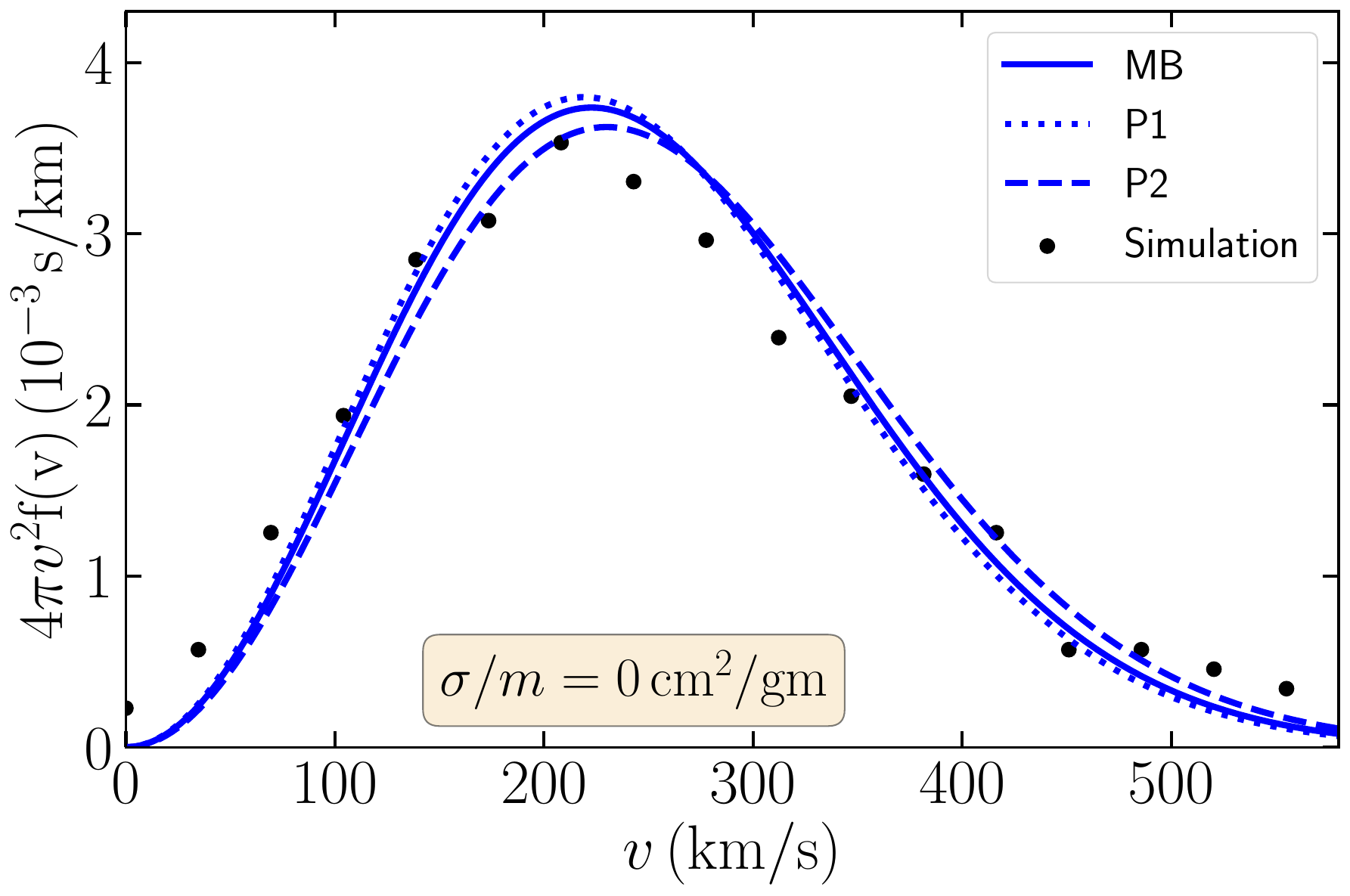}}
		\subfloat[\label{sf:Sirest1}]{\includegraphics[scale=0.191]{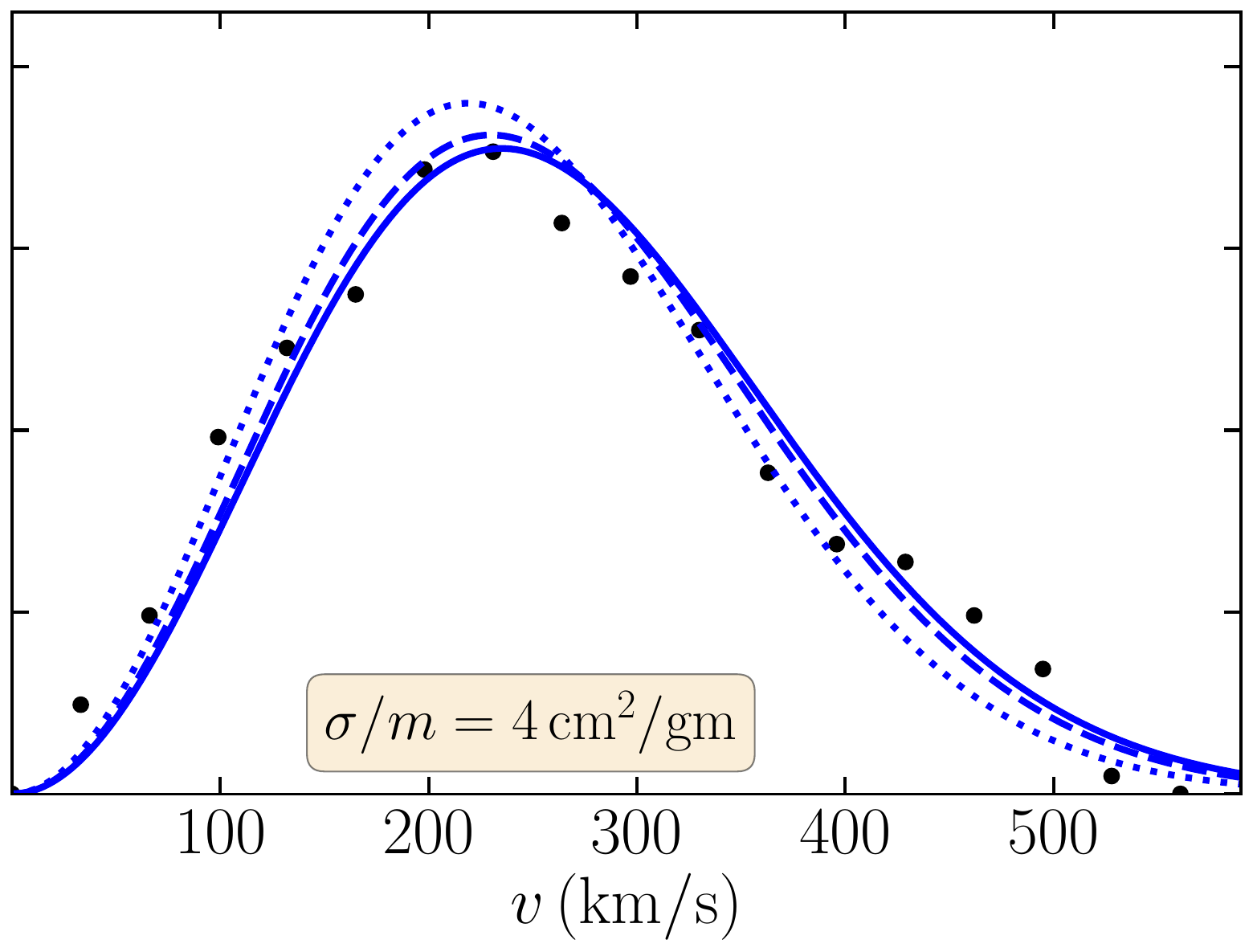}}
		\subfloat[\label{sf:Gerest1}]{\includegraphics[scale=0.191]{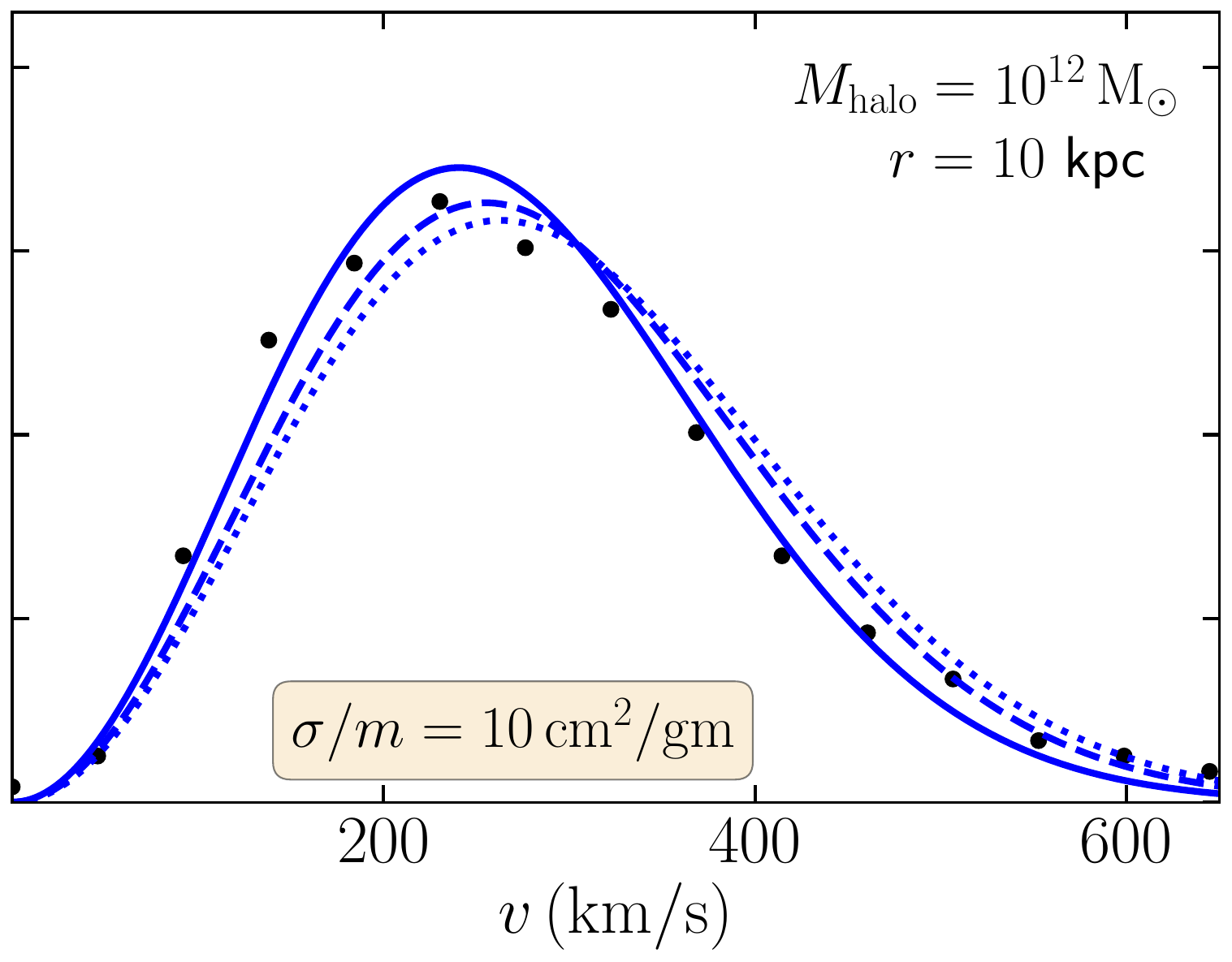}}
		\caption{Fits to the simulated DM velocity distribution in black, for a halo mass of $10^{12}M_{\odot}$. The blue solid curves represent the best fits for the Maxwell-Boltzmann. Dashed and dotted curves represent the Power law and Polynomial profiles associated with the change in MB distribution respectively. The left, middle and right panels are for $\sigma/m$ $0,\,4$ and $10 \, \rm cm^2/gm $ respectively.}
		\label{fig:mwfit}
	\end{center}
\end{figure*}

\begin{figure*}[t]
	\begin{center}
		\subfloat[\label{sf:v0mw1}]{\includegraphics[scale=0.25]{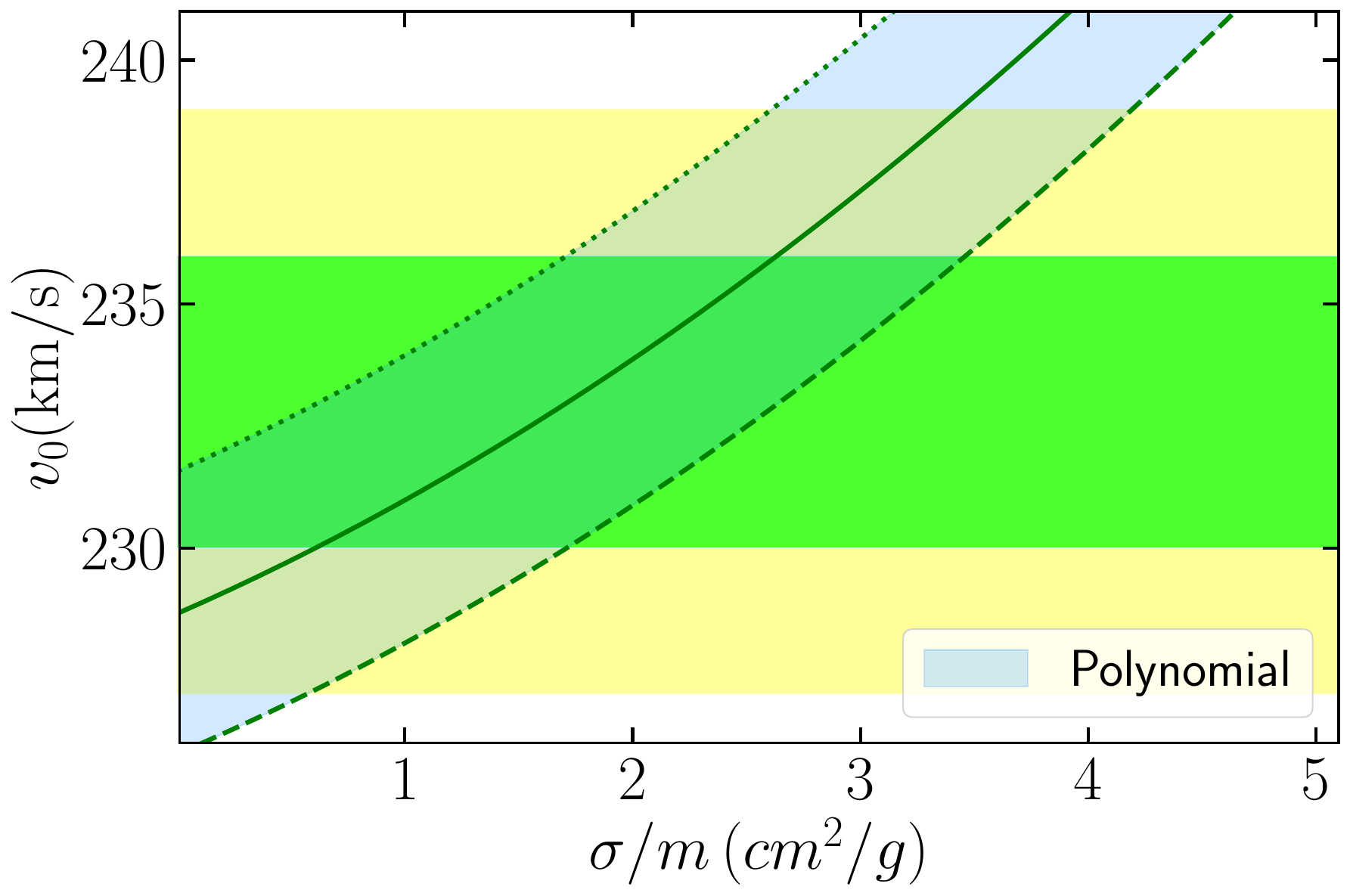}}
		\subfloat[\label{sf:v0mw2}]{\includegraphics[scale=0.25]{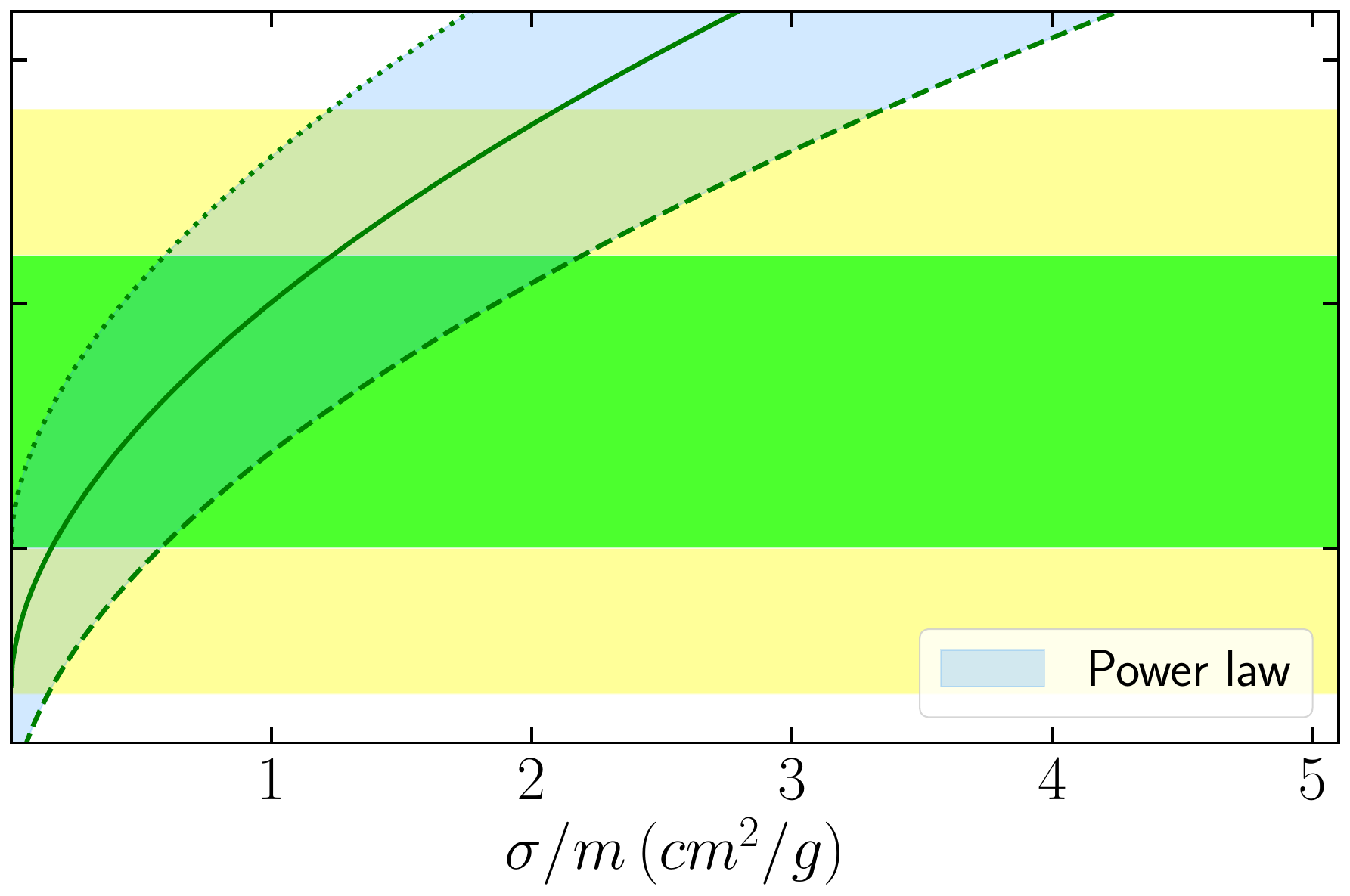}}
		\caption{Variation of the most probable DM velocity, $v_{0}$, as a function of $\sigma/m$ for the benchmark halo mass of Milky-Way. The polynomial and power-law functions are represented in the left and right panel respectively.}
		\label{fig:v0mw}
	\end{center}
\end{figure*}

Given the large uncertainty in the estimated mass of MW like galaxy \cite{eilers2019circular,zhou2023circular,2024MNRAS.528..693O}, we use the set of our $N$-body simulations with halo masses in the range $5\times 10^{11} M_{\odot}$ to $ 5\times10^{12} M_{\odot}$, to mimic a MW size halo. The details for these simulations are given in table \ref{tab:halo}. For a spiral galaxy such as the MW, we study the solar neighborhood as a benchmark position to evaluate the DM velocity distribution and constrain the DM self-scattering cross-section. The solar distance from the galactic center lies within the estimated critical radius within which DM self-interaction is expected to be active \cite{Kaplinghat:2015aga,Sarkar:2022dud} and thus providing a rationale for the assumption of a thermalised MB distribution. The $v_0(R_{\odot})$ is the most probable velocity of the assumed MB distribution of a virialized DM halo, typically measured in the galactic frame. However, it is often approximated with the rotational velocity of the local star, the Sun, measured in the local standard of rest (LSR) frame $(v_0(R_\odot) \sim v_\odot).$ This approximation underlines the fact that the almost collision-less stars inside a galaxy acts as tracers for the DM halo dynamics \cite{Binney:1987bin}.
\begin{table}[t]
	\centering   
	\bigskip
	\resizebox{15.5cm}{!}{%
	\begin{tabular}{|p{0.1cm}|p{0.3cm}|p{1.8cm}|p{1.9cm}|p{1.8cm}|p{1.9cm}|p{1.8cm}|p{1.9cm}|}
		\hline 
	    \multicolumn{1}{|c}{Model} & & \multicolumn{1}{c}{Milky-Way } & & \multicolumn{1}{c}{LSB} & & \multicolumn{1}{c}{Cluster} &  \\ 
	\hline \hline
	        \multicolumn{1}{|c}{}& & Prior& Derived & Prior & Derived & Prior & Derived \\ 
	\hline \hline
			& $a_0$ & $[100,250]$ & $204.2\pm 2.36$ & $[20,80]$ & $47.9 \pm 3.01$ & $[200,900]$ & $512.8\pm12$ \\ 
			& $a_1$ & $[0,20]$ & $2.04\pm0.04$ & $[0,10]$& $0.64\pm0.55$ & $[0,30]$ & $2.5\pm1.4$ \\ 
	P1		& $a_2$ & $[0,20]$ & $0.26\pm 0.01$& $[0,10]$&$ 4.45\pm0.25$ & $[0,30]$ & $5\times10^{-3}$ \\ 
			& $b_1$ & $[0,20]$& $0.28\pm 0.01$& $[0,10]$ & $0.08\pm0.03$ & $[0,30]$ & $1.25\pm0.4$ \\ 
			& $b_2$ & $[10^{-5},10^{-2}]$ & $10^{-4}$ & $[10^{-3},10^{-1}]$ & $0.03\pm0.05$ & $[10^{-9},10^{-4}]$ & $10^{-8}$ \\
	\hline \hline		
			& $a_0$ & $[100,250]$ & $124\pm2.89$ & $[20,80]$ & $45.0\pm 3.0$ & $[200,900]$ & $213.7\pm10$ \\ 
			& $a$   & $[0,20]$ & $7.85\pm0.09$ & $[0,10]$ & $4.45\pm1.1$ & $[0,30]$ & $20.9\pm 0.9$ \\
	P2		& $\alpha$& $[0,2]$ & $0.55\pm 0.01$ & $[0,2]$& $0.54\pm0.2$ & $[0,2]$ & $0.6\pm 0.2$ \\ 
			& $b$  & $[0,20]$ & $16.98\pm 0.19$ & $[0,10]$ & $4.95\pm1.0$ & $[0,30]$ & $7.7 \pm 0.2$ \\ 
		    &$\beta$ & $[0,2]$ & $0.40\pm 0.01$ &  $[0,2]$  &$1.10\pm0.1$ & $[0,2]$ & $0.42\pm 0.01$ \\ 
		\hline 
	\end{tabular}}
	\caption{Here we quote the range of priors used in our MCMC analysis for the three different classes of galaxies used in this work. P1 and P2 correspond to the polynomial and power-law functional forms of $v_0$ respectively. We also quote the numbers used in deriving the bounds, that are shown in figures \ref{fig:v0mw},\ref{fig:v0lsb} and \ref{fig:v0clt}.}
	\label{tab:mwsim}
\end{table}
Recent estimates, using the apparent proper motion of Sgr $A^*$, relative to a distant quasar \cite{bland2016galaxy} and the results of GRAVITY collaboration, which estimate the value of $R_{\odot}$ with reasonably high accuracy \cite{GRAVITY:2018ofz}, have improved the circular velocity of the Sun. The value of $v_0$ at a galactocentric distance of $8.2$ kpc is measured to be $(233\pm 6)\,\rm km/s$ at $95\%$ confidence level \cite{Evans:2018bqy}. This observation sets the initial conditions and range of priors for our MCMC sampler, to obtain posteriors on the parameter space, for the two profiles $P1$ and $P2$, using the set of $N$-body simulation of SIDM halos. The posterior probability distribution functions (PDFs) are given in figure \ref{fig:mwpdf1}, demonstrating a relatively acceptable goodness of fit. The relevant numbers are collected in table \ref{tab:mwsim}. The fit of profiles $P1$ and $P2$, to the simulated data for some benchmark scenario is demonstrated in figure \ref{fig:mwfit}.
Finally, we compare the fitted profiles $P1$ and $P2$ set at MW halo mass of $8\times10^{11} M_{\odot}$ with the rotation curve data at the solar position \cite{zhou2023circular} in figure \ref{fig:v0mw}. The polynomial and power-law profiles are shown in the left and right panels respectively. The solid blue curves denote the mean, whereas the shaded band includes the fitting errors at $\pm 1 \sigma.$ The green and yellow bands in figure \ref{fig:v0mw} denote the $1\sigma$ and $2\sigma$ uncertainty around the measured solar rotational velocity of $233\pm 6 \,\rm km/s$. A constraint of $\sigma/m < 2.7(3.5)\,\rm cm^2/gm$ at a $95\%$ C.L, are obtained for the power-law(polynomial) fits, that are comparable to the Bullet Cluster limits on DM self-interaction $\sim 1.25 \, \rm cm^2/gm $ \cite{Markevitch:2003at,Robertson:2016xjh}. In our analysis of the Milky-Way halo, we ignored the baryon contribution to thermalisation \cite{2019JCAP...10..037D}. It has been argued that for our region of interest around the solar neighborhood, baryons have a significant effect on DM, as the light to mass ratio remains fairly large \cite{2023ApJ...946...73Z}. This should further aid thermalization effects, implying our DM only limits are conservative.
\subsection{LSB galaxy F563-1}
\label{subsec:dwarf}
\begin{figure*}[t]
	\begin{center}
		\subfloat[\label{sf:lsbpolymargin1}]{\includegraphics[scale=0.267]{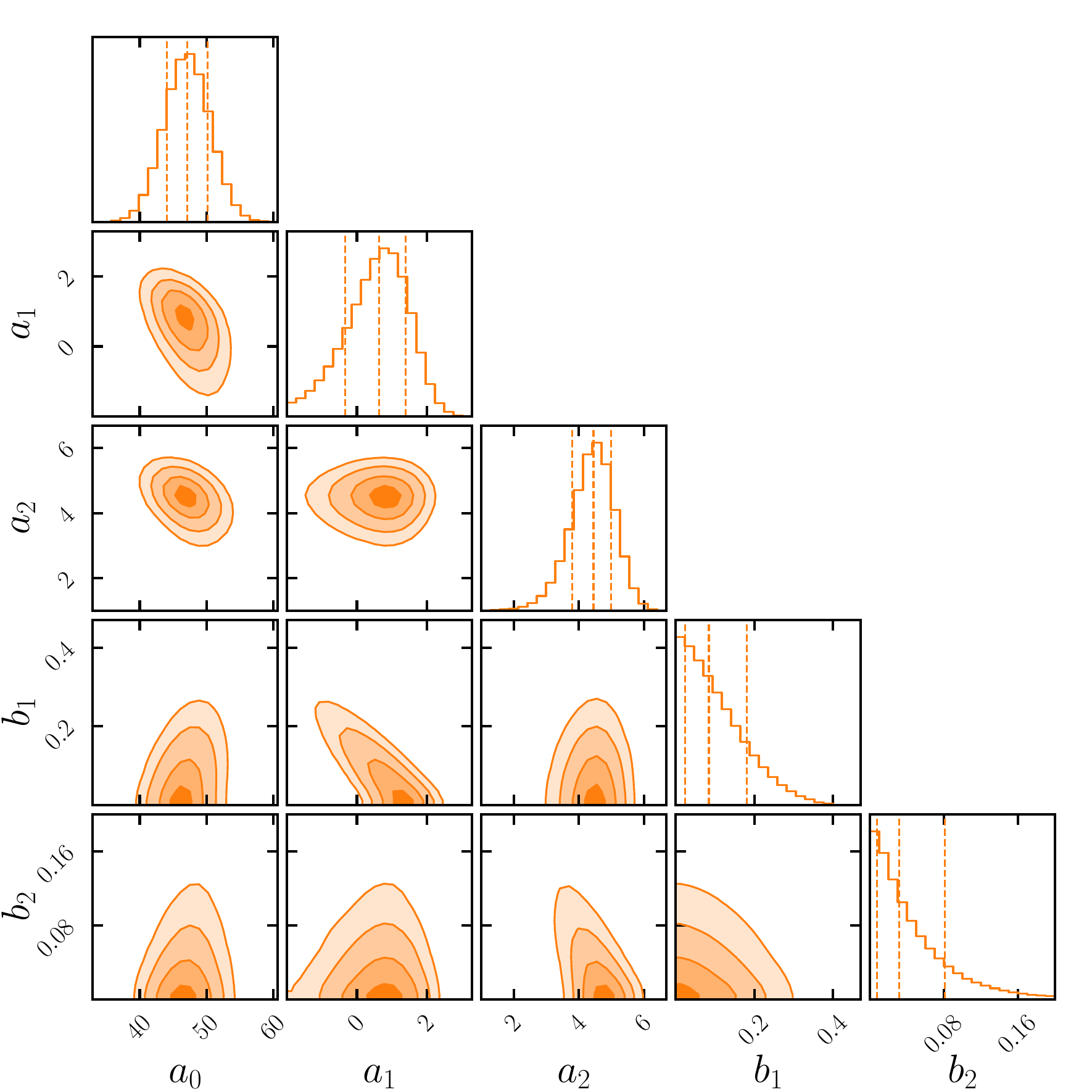}}
		\subfloat[\label{sf:lsbpolymargin2}]{\includegraphics[scale=0.267]{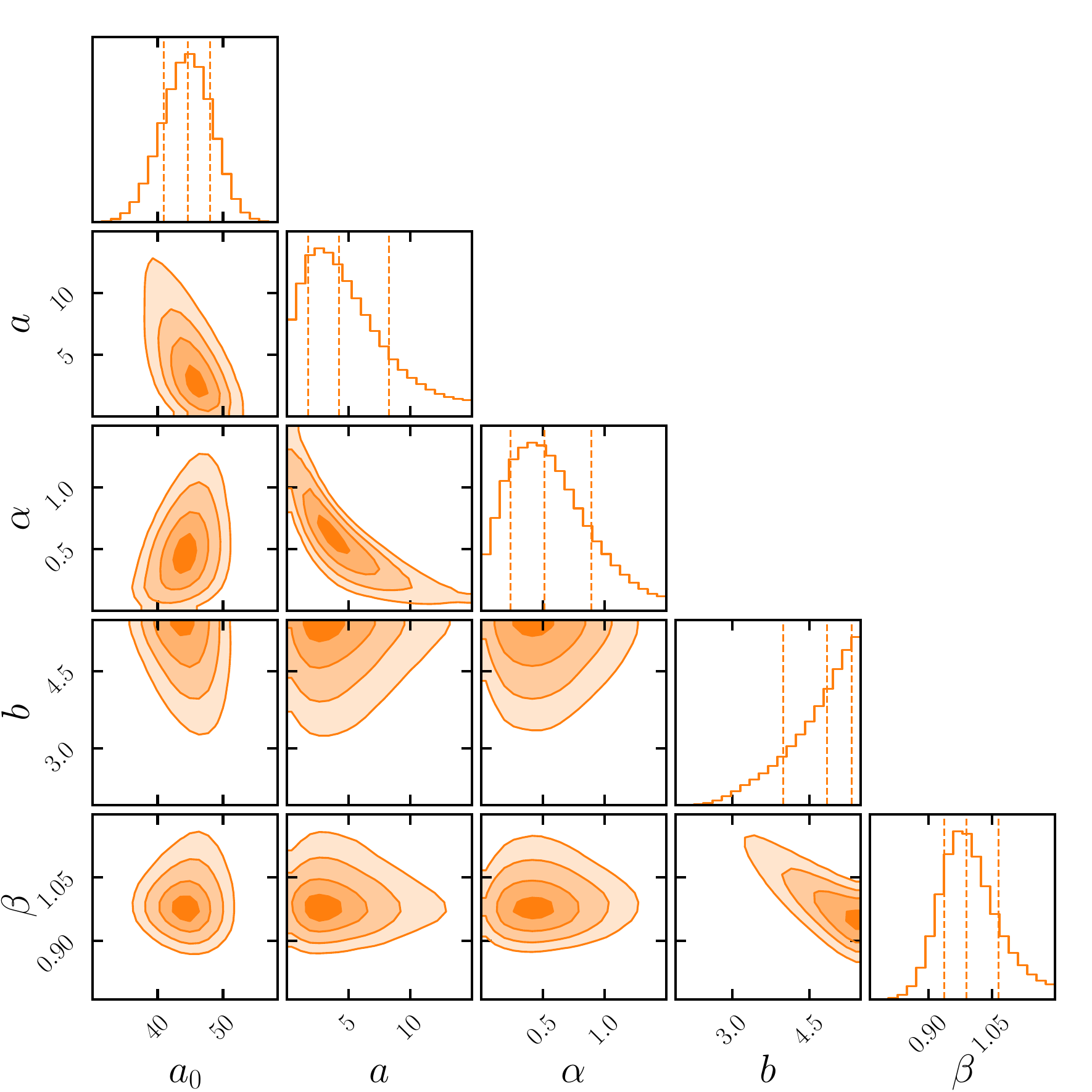}}
		\caption{Posterior PDF for halos in the mass range of LSB galaxies considered. Initialized for the rotation velocity of F563-1. The polynomial and power-law functions are represented in the left and right panel respectively.}
		\label{fig:marginlsb}
	\end{center}
\end{figure*}
\begin{figure*}[t]
	\begin{center}
		\subfloat[\label{sf:Xerest1}]{\includegraphics[scale=0.191]{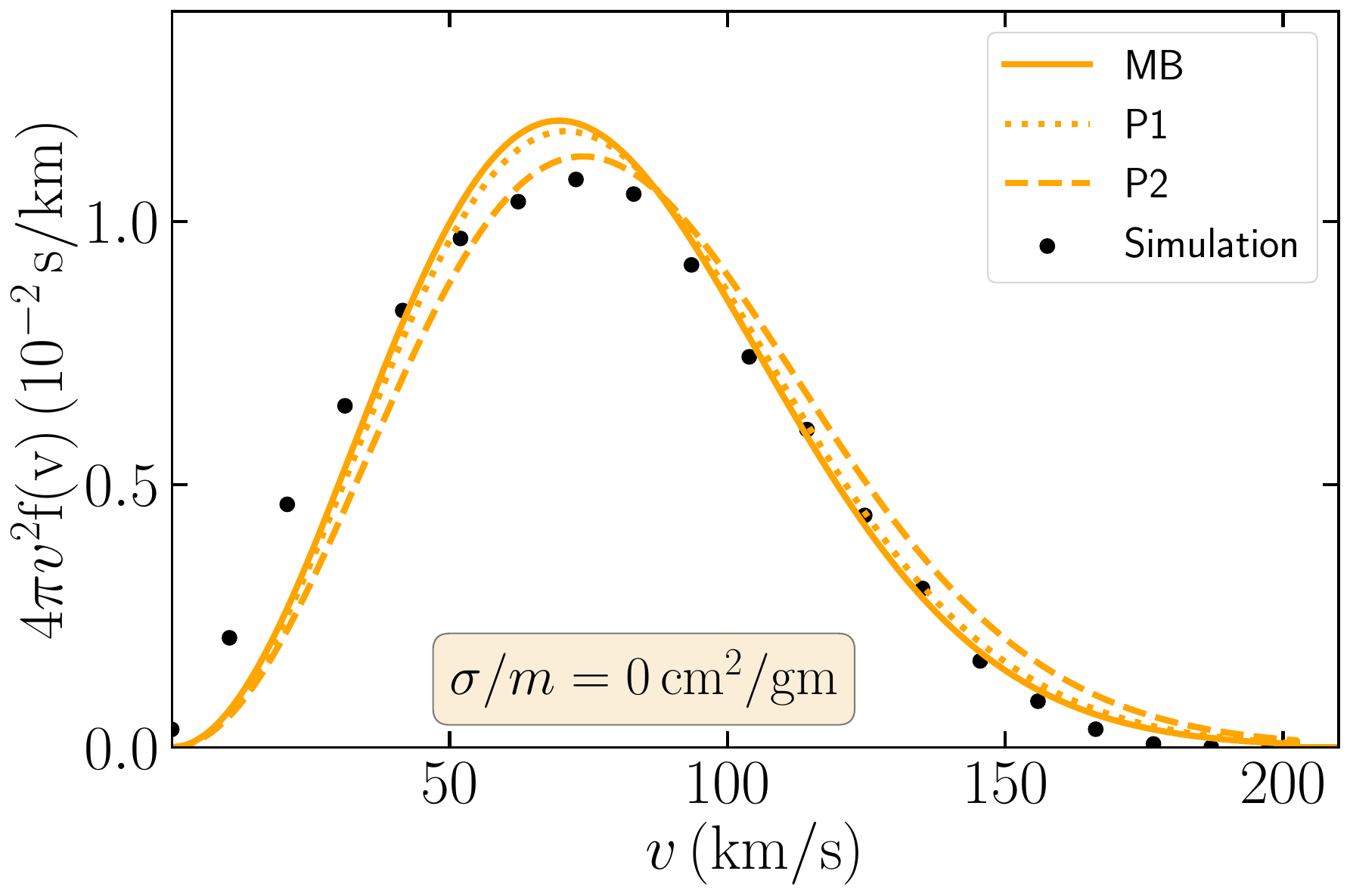}}
		\subfloat[\label{sf:Sirest1}]{\includegraphics[scale=0.191]{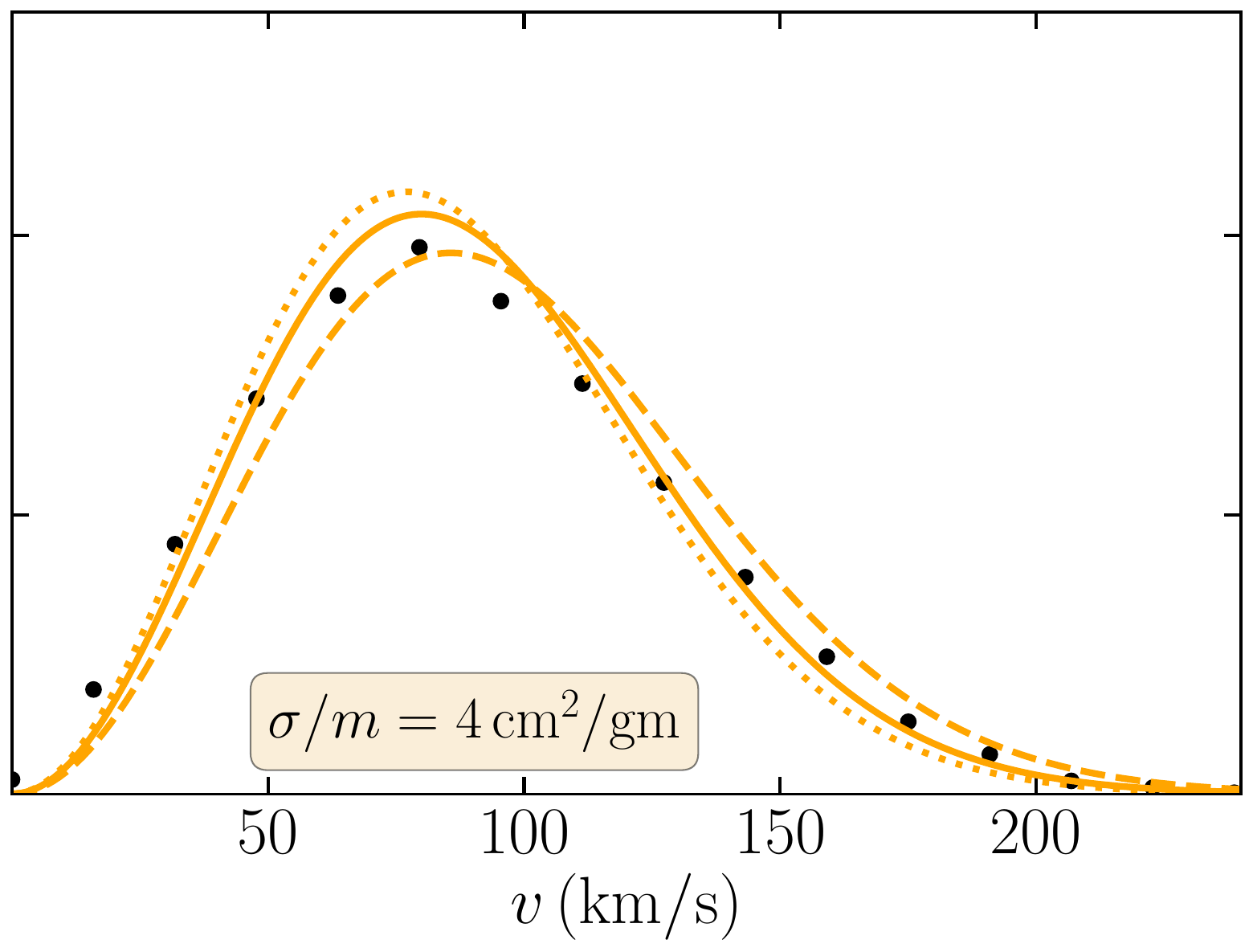}}
		\subfloat[\label{sf:Gerest1}]{\includegraphics[scale=0.191]{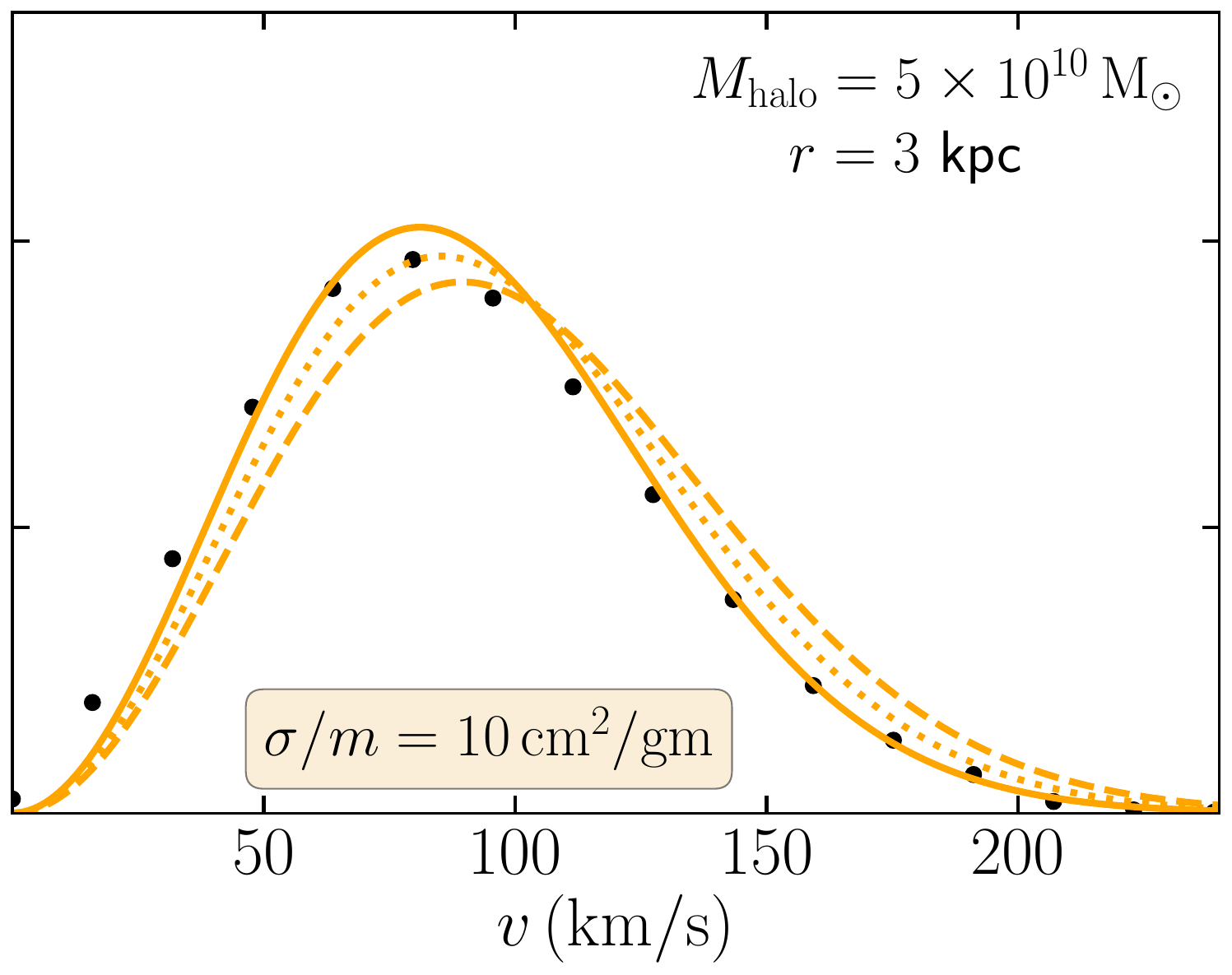}}
		\caption{Same as figure \ref{fig:mwfit}, with the simulated DM velocity distribution for halo mass $5\times 10^{10}M_{\odot}$.}
		\label{fig:lsbfit}
	\end{center}
\end{figure*}
Dwarf and LSB galaxies are ideal for probing DM properties using galactic dynamics, owing to their low baryon contamination \cite{Martin-Alvarez:2022tfw,Roberts:2024uyw,Buckley:2014hja,Akita:2023yga,Wang:2024moc}. We now  consider the effects of DM self-scattering on the velocity distribution in a dwarf galaxy with low gas-cooling rate \cite{Roberts:2024uyw}. Following the procedure described in section \ref{subsec:MW}, we initialize and thereby set the priors for our MCMC analysis from the rotation curve of F563-1, for an observed velocity of $(60.89 \pm 7.76) \,\rm km/s $, at a radius of $2.26$ kpc \cite{KuziodeNaray:2007qi} from the halo center. For mimicking LSB size halos, we select DM halo masses in the range $10^{10}M_{\odot}-10^{11}M_{\odot}$, from our set of $N$-body simulations. (refer table \ref{tab:halo}). The rotation curve data of F563-1 has been taken for the minimum disc model \cite{deBlok:2001hbg,KuziodeNaray:2007qi}, which uses the DensePak integral field spectroscopy \cite{KuziodeNaray:2006wh}. The rotational curve data in the inner regions ($r<3$ kpc) for F563-1 is similar to that of NGC4396 and F568-V1. Hence the results of our analysis is valid and compatible for dwarf and low LSB galaxies having similar circular velocity profiles which are low in baryon content.
\begin{figure*}[t]
	\begin{center}
		\subfloat[\label{sf:v0lsb1}]{\includegraphics[scale=0.25]{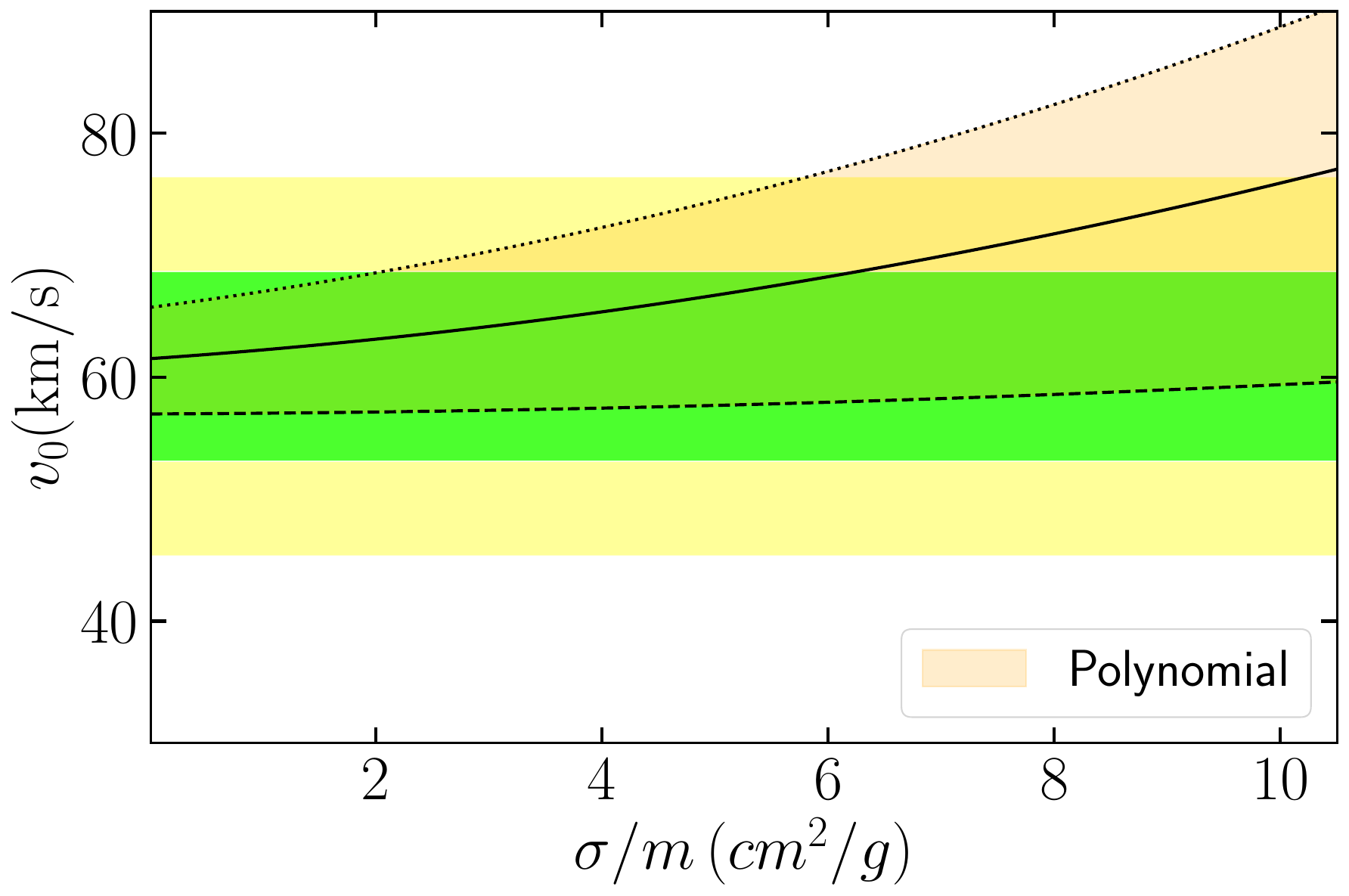}}
		\subfloat[\label{sf:v0lsb2}]{\includegraphics[scale=0.25]{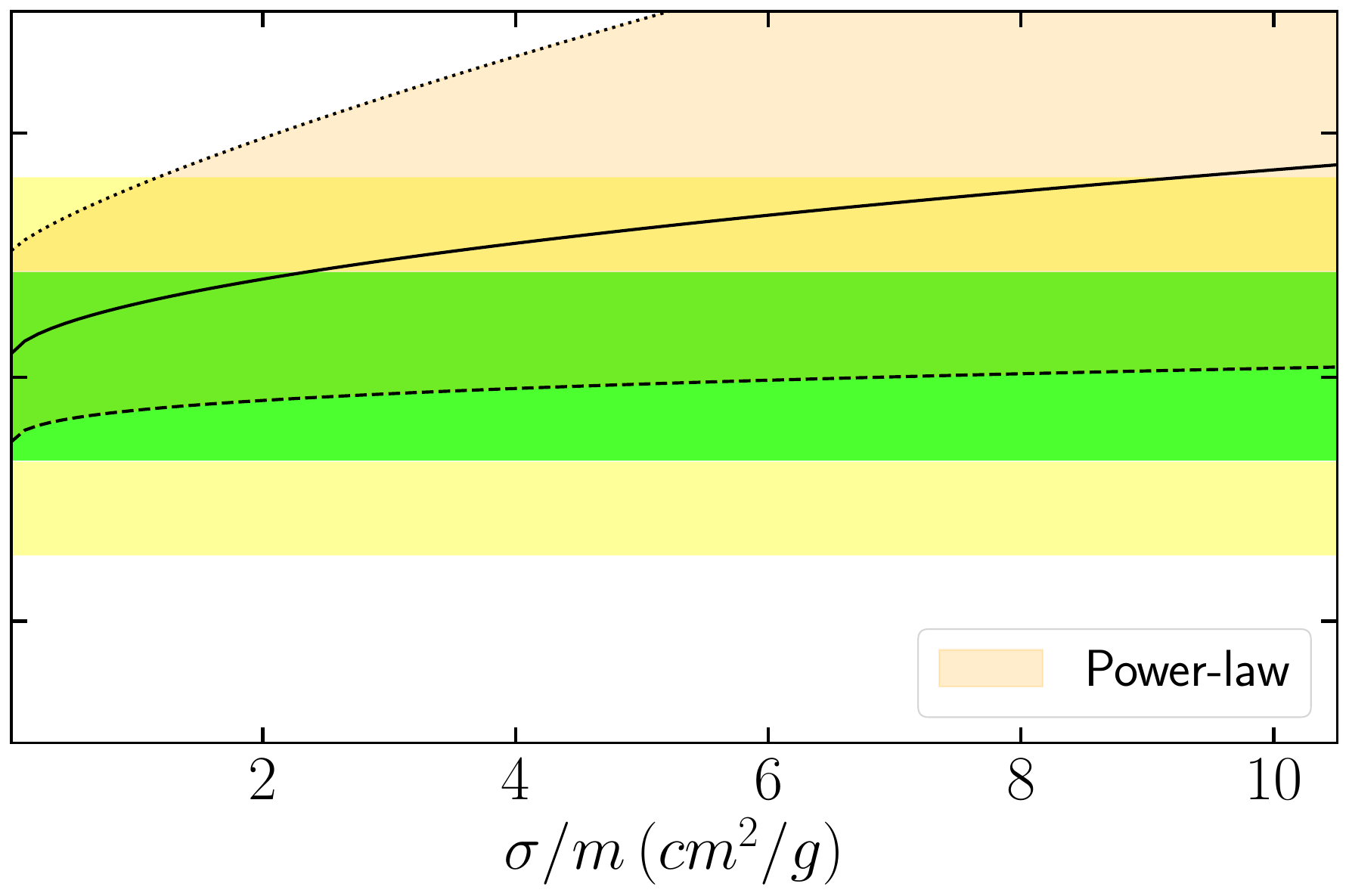}}
		\caption{Variation of the most probable DM velocity $v_{0}$, as a function of $\sigma/m$ for the estimated halo mass of F563-1. The polynomial and power-law functions are represented in the left and right panel respectively.}
		\label{fig:v0lsb}
	\end{center}
\end{figure*}
We show the posterior PDFs in figure \ref{fig:marginlsb} and the numerical values are tabulated in table \ref{tab:mwsim}. In figure \ref{fig:lsbfit}, some benchmark fits are demonstrated by comparing the fitted profiles $P1$ and $P2$ with the SIDM $N$-body simulated data. We report a conservative bound $<9.8\, \rm cm^2/gm$ in figure \ref{fig:v0lsb} comparing the fitted profiles with the observed rotation velocity at $2.26$ kpc from the galactic center for F563-1. For F563-1, the bounds derived in this method are weaker in comparison to the complementary ones derived from classic dwarf spheroidal \cite{Ebisu:2021bjh} and ultra-faint dwarf galaxies associated with the Milky-Way, which take into the effect self-interactions with \cite{Ando:2025qtz} and without core collapse \cite{Hayashi:2020syu}. However, constraints on MW sub-halos from semi-analytic studies using simulations and observations predict $\sigma/m \sim (3-10) \, \rm cm^2/gm $ \cite{Valli:2017ktb}. This indicates that a larger statistics from box simulations may have noticeable effects on the bounds presented in this study.
\subsection{Galaxy cluster A611}
\label{subsec:cluster}
\begin{figure*}[t]
	\begin{center}
		\subfloat[\label{sf:Xerest1}]{\includegraphics[scale=0.267]{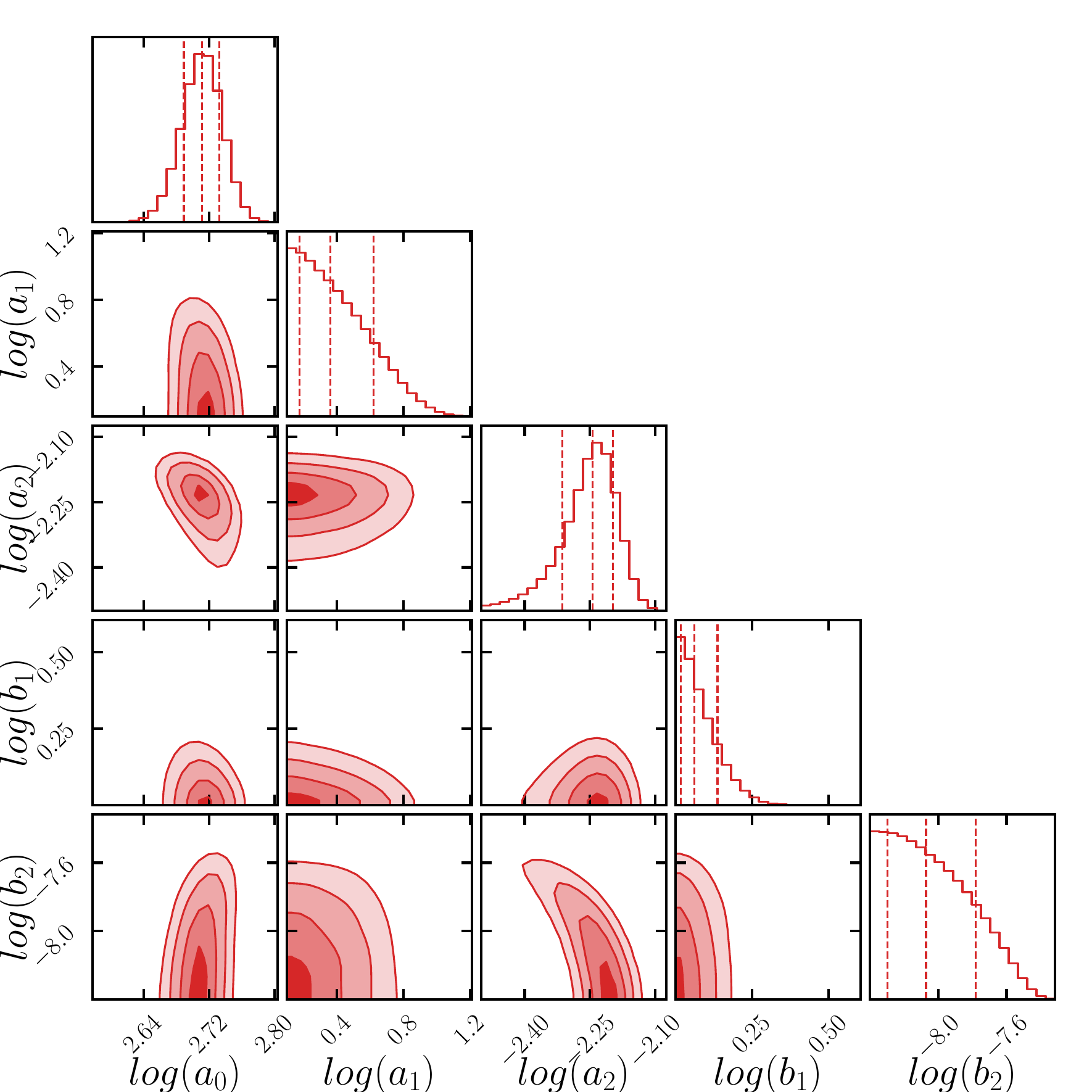}}
		\subfloat[\label{sf:Xerest1}]{\includegraphics[scale=0.267]{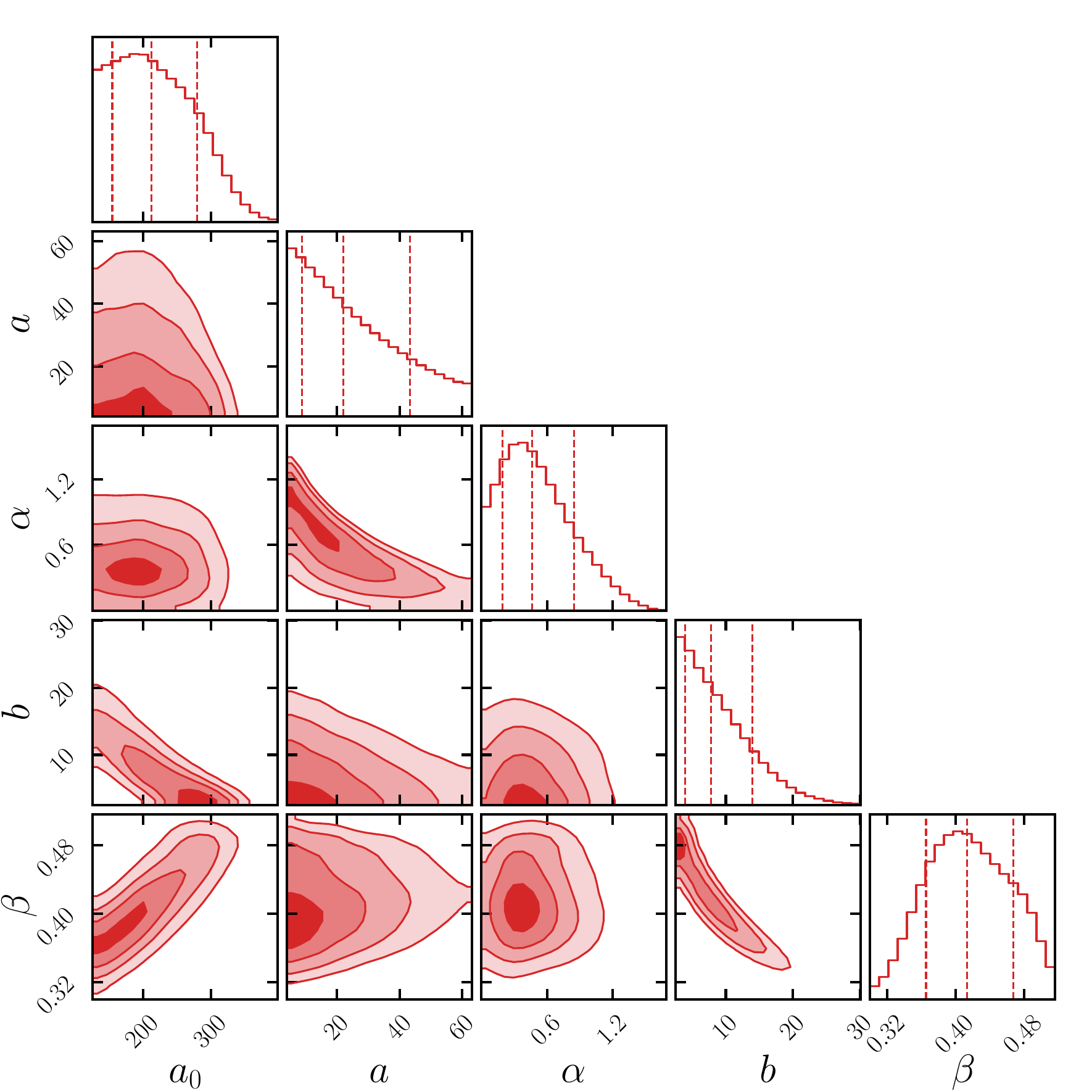}}
		\caption{Posterior PDF for halos in the mass range of galaxy clusters considered. Initialized for the A611 cluster. The polynomial and power-law functions are represented in the left and right panel respectively.}
		\label{fig:cltpdf}
	\end{center}
\end{figure*}
\begin{figure*}[t]
	\begin{center}
		\subfloat[\label{sf:Xerest1}]{\includegraphics[scale=0.191]{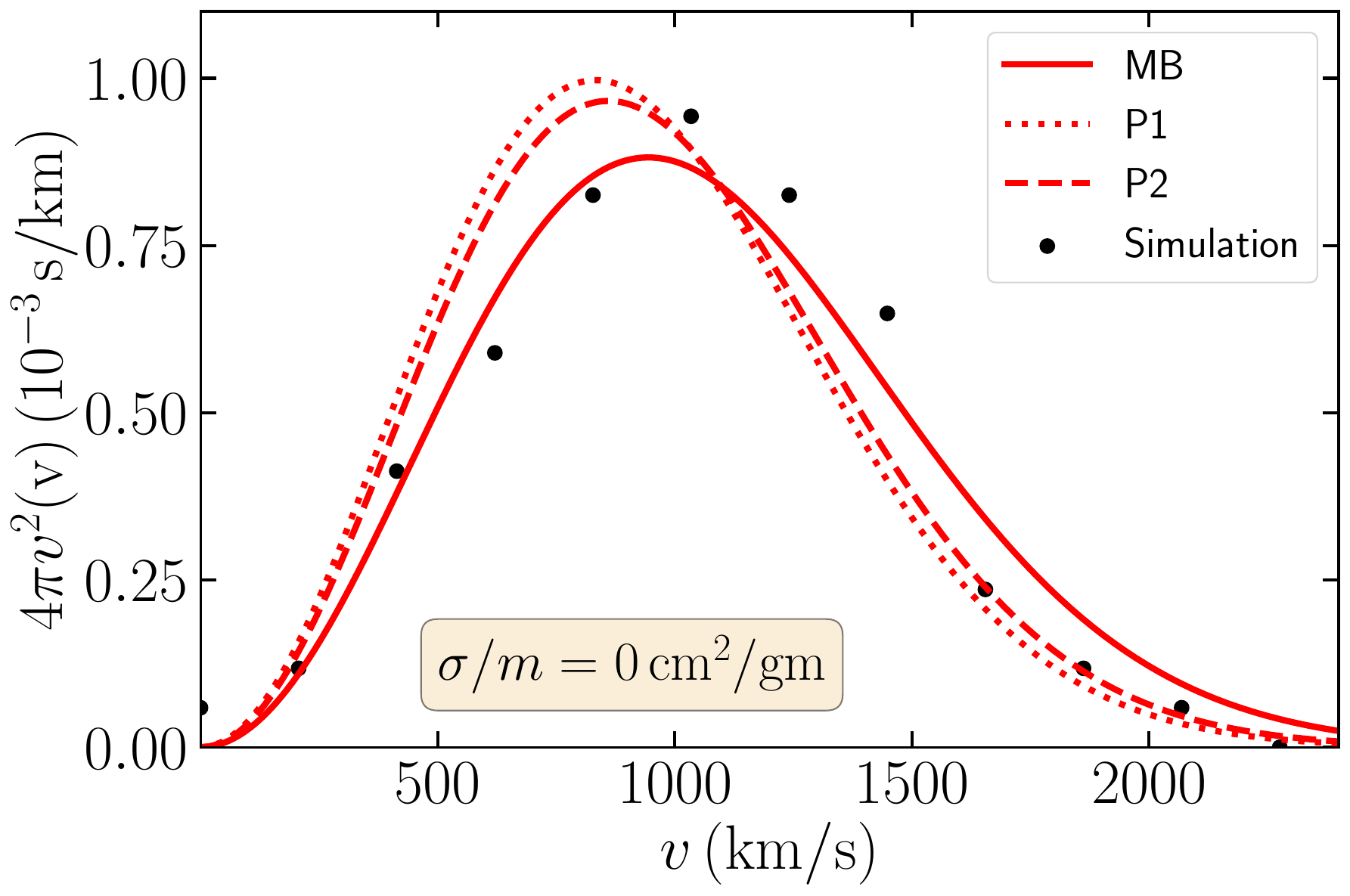}}
		\subfloat[\label{sf:Sirest1}]{\includegraphics[scale=0.191]{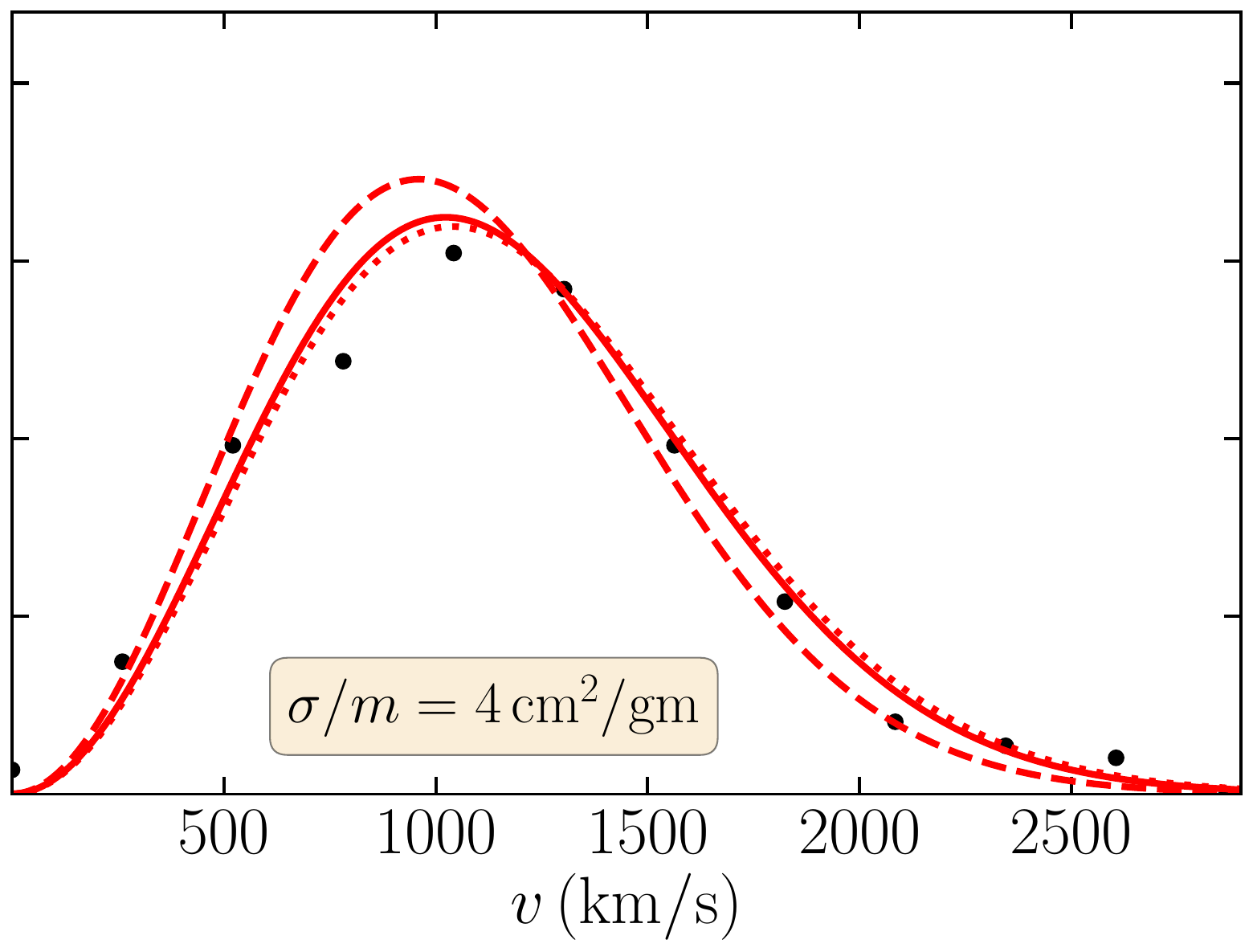}}
		\subfloat[\label{sf:Gerest1}]{\includegraphics[scale=0.191]{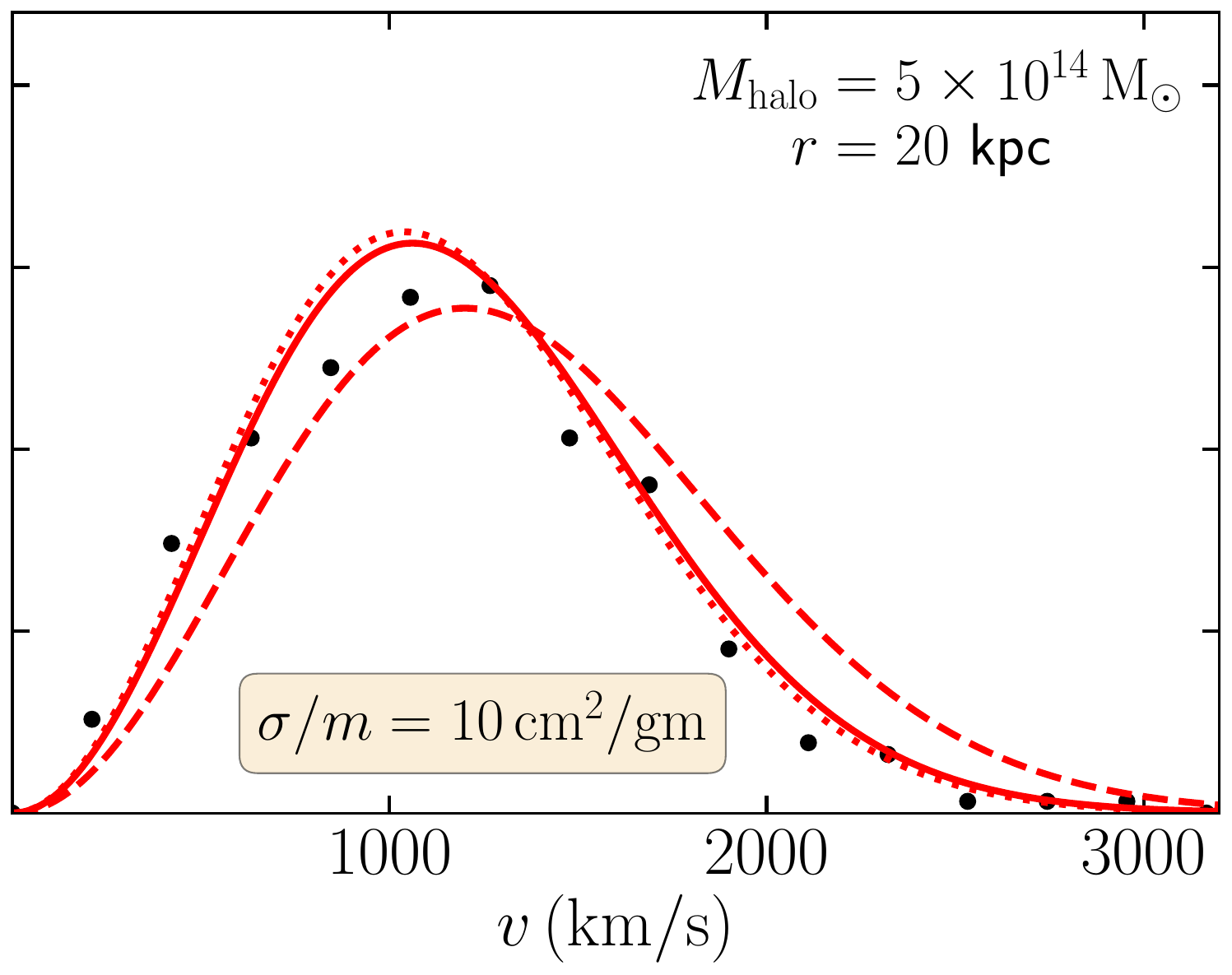}}
		\caption{same as figure \ref{fig:mwfit}, for halo mass $5\times 10^{14}M_{\odot}$.}
		\label{fig:clustfit}
	\end{center}
\end{figure*}
\begin{figure*}[t]
	\begin{center}
		\subfloat[\label{sf:v0clt1}]{\includegraphics[scale=0.25]{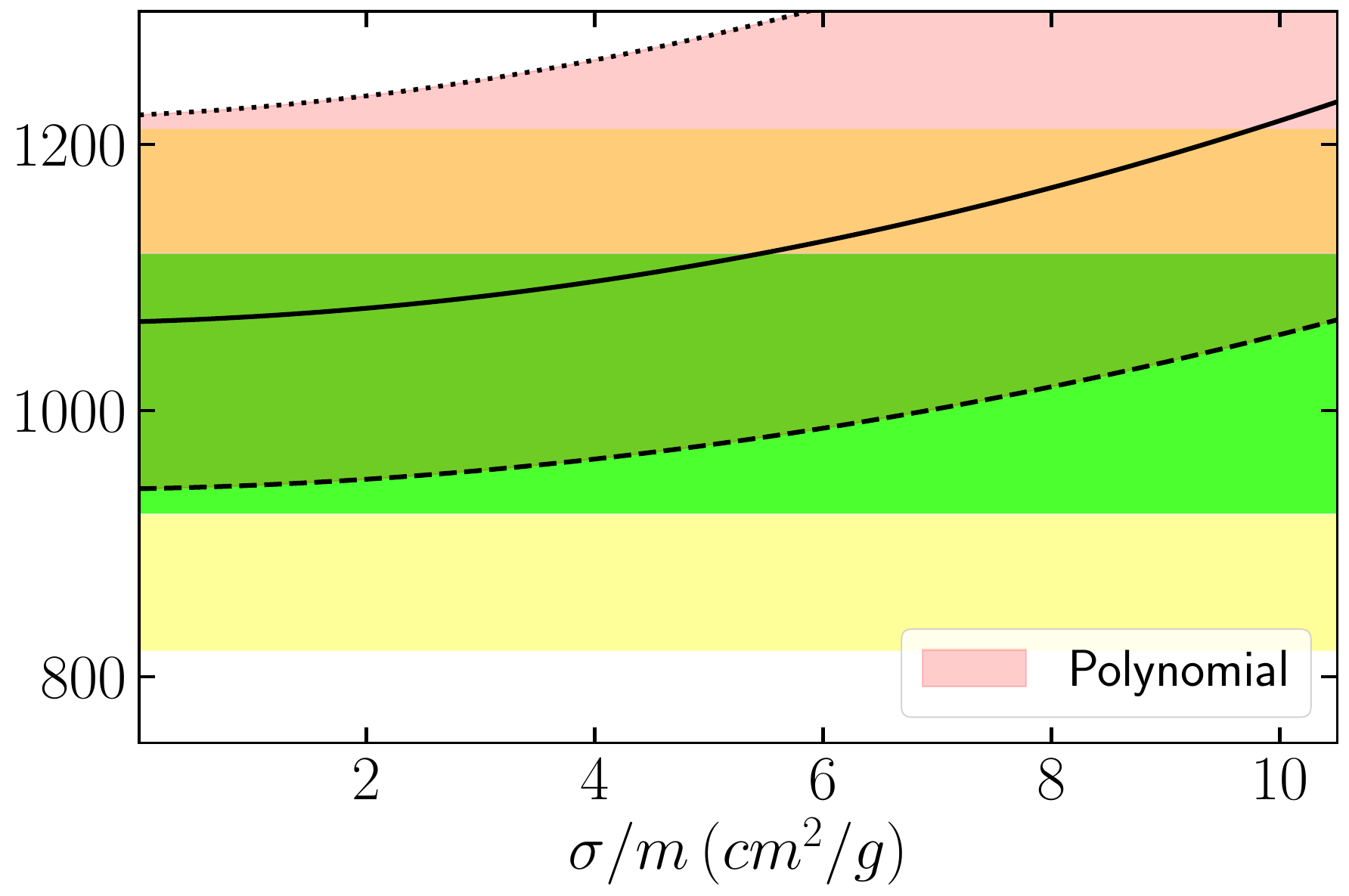}}
		\subfloat[\label{sf:v0clt2}]{\includegraphics[scale=0.25]{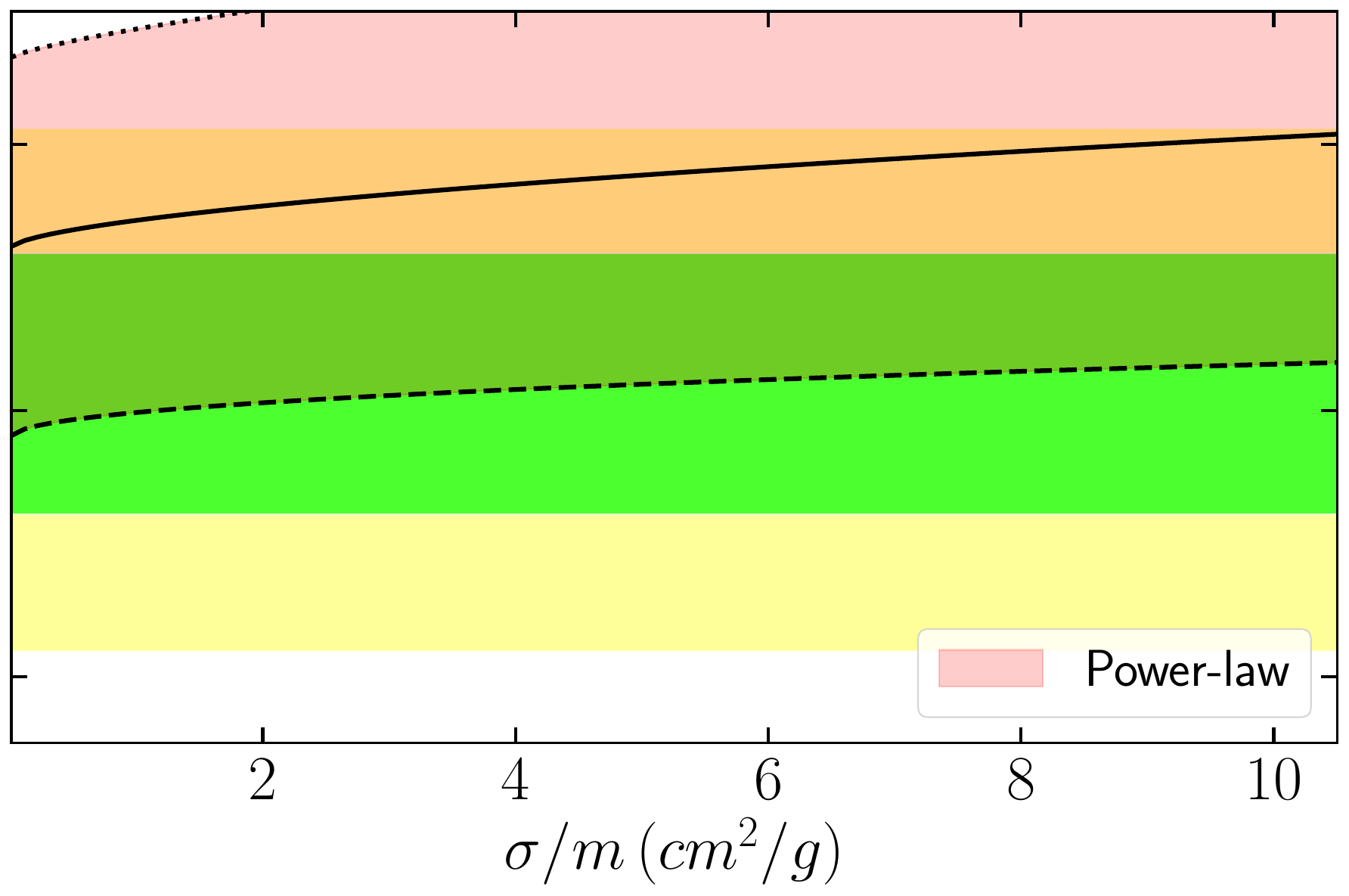}}
		\caption{Variation of the most probable DM velocity given by $v_{0}$, as a function of $\sigma/m$ for the estimated halo mass of A611. The polynomial and power-law functions are represented in the left and right panel respectively.}
		\label{fig:v0clt}
	\end{center}
\end{figure*}

Some of the initial constraints on DM self-interactions have been inferred from the offset between  X-ray and gravitational lensing of stars and DM within clusters. The Bullet cluster \cite{Randall:2007ph,Robertson:2016xjh}, baby bullet \cite{Bradac:2008eu}, Abell \cite{Kahlhoefer:2015vua}, and others \cite{10.1046/j.1365-8711.2001.04477.x,Banerjee:2019bjp,Sabarish:2023ija,Sirks:2024njj} provide some of the well known constraints in literature. We use the  Abell 611 cluster (A611), located at a redshift of 0.288, for constraining $\sigma/m$ of SIDM from velocity distribution profiles. The kinematics of A611 has been studied in \cite{New:2012and} with approximately 236 galaxies, having low peculiar velocities. Using the data for resolved stellar velocity dispersion$ (\sigma_o)$ and assuming a Maxwell-Boltzmann mean velocity of $\left< v\right> \sim 2\sigma_o$ \cite{Kaplinghat:2015aga}, we work with $\left< v\right> \sim \rm 993\rm \, \pm98.15\, km/s$, at an approximate radial distance of $\sim 20$ kpc from the BCG of A611. We use this value to set priors and initialize our MCMC analysis for our simulated halos in the mass range of $10^{13}\,M_{\odot}$ to $10^{15}\,M_{\odot}$ (refer table \ref{tab:halo}). We plot the posterior PDFs\footnote{Logarithmic ranges are used for the polynomial form to find better convergence. The corresponding column in table \ref{tab:mwsim} has been scaled accordingly.} in figure \ref{fig:cltpdf}, and tabulate the priors and posteriors in table \ref{tab:mwsim}. In figure \ref{fig:clustfit} we plot the representative distributions, comparing the fitted profiles, P1 and P2 with the simulated data. Ignoring the sub-structure of these clusters \cite{Zhang:2024fib}, we read off from figure \ref{fig:v0clt} a conservative bound $\sigma/m \sim 10\,\rm cm^2/gm$ for both the profiles at $95\%$ confidence. This number although large, lies in the ballpark of other bounds derived from clusters \cite{Harvey:2015hha}. 
In Figures \ref{fig:v0mw}, \ref{fig:v0lsb} and \ref{fig:v0clt}, the shaded bands denote the uncertainty associated with our analysis primarily arising from fits to the simulated data. As the Monte-Carlo chain has been initialized with a collision-less DM scenario, we notice an increase in $v_0$ with the scattering cross-section. For Milky-way sized halos, both profiles mimic the simulated data reasonably well. However, the F563-1 LSB galaxy seems to prefer the power-law function. Fits worsen for the cluster, evident from figures \ref{fig:clustfit} and \ref{fig:v0clt}.
\section{Conclusion}
\label{sec:conc}
It has been well documented in the literature that dark matter self-interactions and baryon dynamics are competing factors that can contribute to the modifications of structures at galactic scales. This can have impact on small scale structure issues like the formation of galactic core and the related diversity issues with rotation curves, formation of satellite galaxies, sub-halos or stellar streams. In this work we utilize the observation of galactic rotation curves to constrain dark matter self-interaction from its impact on the velocity distribution function of isolated virialized dark matter halos.

We consider elastic and isotropic self-interactions of a single component DM as a contributing mechanism for redistribution of energy within a galaxy, thus impacting their thermal velocity distribution. We perform an extensive SIDM only $N$-body simulations of isolated halos ranging between LSB galaxy to spiral galaxy to galaxy cluster, incorporating velocity independent self-interaction between DM particles. Our simulations indicate that the self-interactions effectively contribute to thermalise the dark mater within the core region where the interaction rate is numerically significant. 

Our approach is to capture the leading effect of self-interaction inside the core region of a SIDM halo, in terms of modification of the most probable velocity of MB distribution. This distortion in the most probable velocity beyond its  usual  mapping to local stellar rotational velocity is expected to happen due to energy redistribution in the halo, facilitated by the dark matter self-interactions along with baryonic effects. We compare these distortions captured in our SIDM $N$-body simulations with the observational errors in the measurement of the rotational velocity of stars in specific galaxies. This provides a handle to constrain the self-interaction of the dark matter that is comparable to existing bounds. We report conservative bound of $\sigma/m\leq2.7\,\rm cm^2/gm$ at $95\%$ confidence from the solar neighborhood of MW like galaxies. For LSB galaxies with inner rotation curves similar to F563-1, we report a bound of $\sigma/m\leq9.8\,\rm cm^2/gm$. Whereas, for a galaxy cluster such as A611 we obtain a more relaxed  constraint at $\sim10\,\rm cm^2/gm$.
\acknowledgments
The authors would like to thank Somnath Bharadwaj for discussions during the initial part of this work. S.S. thanks Tarak Nath Maity, Alejandro Ibarra and Hai-Bo Yu for various helpful comments. We acknowledge Abinash Kumar Shaw for discussions on MCMC analysis. This work used the Supercomputing facility of IIT Kharagpur established under National Supercomputing Mission (NSM), Government of India and supported by Centre for Development of Advanced Computing (CDAC), Pune. S.S. is supported by the Basic Research Laboratory Program of the National Research Foundation of Korea (Grant No. RS-2022-NR070815).
\FloatBarrier
\bibliographystyle{JHEP}
\bibliography{interactivedm.bib}

\end{document}